\documentclass[twocolumn,aps]{revtex4}
\usepackage[utf8x]{inputenc}
\usepackage{ucs}
\usepackage{graphicx}
\usepackage{epsfig}
\usepackage{amsmath, amsthm, amssymb}
\usepackage{amsfonts}
\usepackage{ucshyper}
\usepackage{fancyhdr}

\def\nim{Nucl. Instr. and Meth.~}

\def\etal{{\it et al.~}}
\def\eg{{\it e.g.,~}}
\def\ie{{\it i.e.~}}

\def\vs{{\it vs.~}}
\newcommand{\TheAuthor}{}

\newcommand{\TheTitle}{}

\makeindex

\begin{document}

\title{DUAL-READOUT CALORIMETRY}

\author{Sehwook Lee}
\email{sehwook.lee@knu.ac.kr}
\affiliation{Department of Physics, Kyungpook National University, Daegu, Korea}
\author{Michele Livan}
\email{michele.livan@unipv.it}
\affiliation{Dipartimento di Fisica, Universit\`a di Pavia and INFN Sezione di Pavia, Via Bassi 6, Pavia 27100 Italy}
\author{Richard Wigmans}
\email{wigmans@ttu.edu}
\affiliation{Department of Physics and Astronomy, Texas Tech University, Lubbock, TX 79409-1051, U.S.A.}
\begin{abstract}
In the past 20 years, dual-readout calorimetry has emerged as a technique for measuring the properties of high-energy hadrons and hadron jets that offers considerable advantages compared with the instruments that are currently used for this purpose in experiments at the high-energy frontier. In this paper, we review the status of this experimental technique and the challenges faced for its further development
\end{abstract}
\maketitle
\tableofcontents

\section{Introduction}
\vskip 2mm
Progress in our understanding of the structure of matter and its fundamental properties has always been driven
by the availability of new, more powerful particle accelerators.
In the past fifteen years, discussions about accelerator projects in the post-LHC era have mainly concentrated on 
a high-energy electron-positron collider, with a center-of-mass energy that would allow this machine to become a 
factory for $t\bar{t}$ and Higgs boson production. Both linear colliders (ILC \cite{ILC}, CLIC \cite{CLIC}) and circular ones (FCCee, CEPC \cite{CEPC}) have been proposed
in this context. A sufficiently large circular collider could subsequently be used to further push the energy frontier for hadron collisions beyond the LHC limits.

In order to take full advantage of the experimental opportunities created by  such colliders, adequate
particle detectors will be needed, since the quality of
the scientific information that can be obtained will, to a very large extent, be determined (and limited) by the quality of the detectors with which experiments at these machines will be performed.
In these experiments, that quality primarily concerns the precision with which the four-vectors of the scattered objects produced in the collisions can be measured. At the TeV scale, these objects are leptons, photons and fragmenting quarks, di-quarks and gluons. The fragmenting objects are commonly referred to as jets. Achieving the best possible precision for the momentum and energy measurements of these objects is usually a very (if not the most) important design goal of the proposed experiments.

These considerations have determined the directions in which calorimeter R\&D has evolved in the past fifteen years.
Two different approaches have been followed:
\begin{enumerate}
\item Particle Flow Analysis
\item Dual-Readout Calorimetry
\end{enumerate}
In the first approach, the calorimeter information is supplemented by data provided by a magnetic tracker for the measurement of jets.
The tracker is used to measure the momenta of the charged jet fragments, while the calorimeter data are used to measure the energy of the neutral ones. Since the charged fragments also develop showers in the calorimeter, the main problem in this approach is {\sl double counting}. The proponents believe that a fine-grained calorimeter structure is the key to solving this problem. The achievements made in this type of R\&D have recently been reviewed by Sefkow and coworkers \cite{Sef15}.

In this paper, we discuss the second approach, which aims at developing a calorimeter system that will allow measurements with excellent precision of all the fundamental objects listed above, even in stand-alone mode.
The discussion focuses on hadron calorimetry, where the challenges will be greatest, although the detection of electromagnetic showers also faces problems, especially in finely segmented instruments \cite{Liv17}.
The dual-readout method is based on the use of two different types of signals, which provide complementary information about the showers developing in the calorimeter. The technique makes it possible to avoid/eliminate many of the problems that have traditionally strongly limited the performance of hadron calorimeters.

In Section II, we present a brief overview of the principles that determine and limit important properties of the current generation of calorimeters. The emphasis is on sampling calorimeters, which are by far the most common ones currently used and which are expected to be the only type for future experiments at the TeV scale. The specific features that play a role at the TeV scale are the focus of a separate subsection. In Section III, the ideas that form the basis of dual-readout
calorimetry are described. An early attempt to exploit the availability of complementary signals (in a very thin detector) is the topic of Section IV. 
In Section V, the potential advantages of dual-readout calorimetry in calorimeters that fully contain the showering particles are examined.
In Sections VI and VII, the DREAM and RD52 projects
are described in some detail and the results discussed. In Section VIII, we give our assessment of the current status and necessary further developments of this intriguing experimental technique. 

\section{Calorimetry - a primer}

In particle physics and related fields (cosmic ray studies, astrophysics,...), a calorimeter is a detector in which the particles to be detected are completely absorbed.
The detector provides a signal that is a measure for the energy deposited in the absorption process.
One frequently distinguishes between {\sl homogeneous} and {\sl sampling} calorimeters.
In a homogeneous calorimeter, the entire detector volume is sensitive to the particles and may 
contribute to the generated signals.
In a sampling calorimeter, the functions of
particle absorption and signal generation are exercised by {\sl different} materials, called 
the {\sl passive} and {\sl active medium}, respectively. The passive medium is usually a high-density
material, such as iron, copper, lead or uranium. The active medium generates the light or charge
that forms the basis for the signals from such a calorimeter.

\subsection{Functions and properties of calorimeters}

Calorimeters measure the energy released in the absorption of (sub)nuclear particles that enter them. 
They generate signals that make it possible to quantify that energy. Typically, these signals
provide also other information about the particles, and about the event in which they were produced.
The signals from a properly instrumented absorber may be used to measure the entire four-vector of the particles. 

By analyzing the energy deposit pattern, the direction of the particle can be measured.
The mass of the showering particle can be determined in a variety of ways, \eg from the time structure of the signals, the energy deposit profile, or a comparison of the measured energy and the momentum of the particle. Calorimeters are also used to identify muons and neutrinos. High-energy muons usually deposit only a small fraction of their energy in the calorimeter and produce signals in downstream detectors. Neutrinos typically do not interact at all in the calorimeter. If an energetic neutrino is produced in a colliding-beam experiment, this phenomenon will lead to an imbalance between the energies deposited in any two hemispheres into which a $4\pi$ detector can be split.
Such an imbalance is usually referred to as {\sl missing transverse energy}.

The latter is an example of the {\sl energy flow information} a calorimeter system can provide.
Other examples of such information concern the {\sl total transverse energy} and the production of {\sl hadronic jets}
in the measured events. Since this information is often directly related to the physics goals of the experiment, and since it can be obtained extremely fast, calorimeters usually play a crucial role in the trigger scheme, through which interesting events are selected and retained for further inspection off-line. 

The calorimeter's properties should be commensurate with the role it has to play in the experiment.
Relevant properties in this context are the energy resolution, the depth (which determines the effects of shower leakage), the time resolution and the hermeticity.

\subsection{Electromagnetic calorimeters}

Electromagnetic calorimeters are specifically intended for the detection of energetic electrons and 
$\gamma$s, but produce usually also signals when traversed by other types of particles. They are used over a very wide energy range, from the semiconductor crystals that measure $X$-rays down to a few keV to shower counters that orbit the Earth on satellites in search for electrons, positrons and $\gamma$s with energies $>10$ TeV.
These calorimeters don't need to be very deep, especially when high-$Z$ absorber material is used. For example, when 100 GeV electrons enter a block of lead, $\sim 90\%$ of their energy is deposited in only 4 kg of material.
%
By far the best energy resolutions have been obtained with large semiconductor crystals, and in particular high-purity germanium. These are the detectors of choice in nuclear $\gamma$ ray spectroscopy, and routinely obtain relative energy resolutions ($\sigma/E$) of 0.1\% in the 1 MeV energy range \cite{Lla72}. The next best class of detectors are scintillating crystals, which are often the detectors of choice in experiments involving $\gamma$ rays in the energy range from
1 - 20 GeV, which they measure with energy resolutions of the order of 1\% \cite{Gra94}. Excellent performance in this energy range has also been reported for liquid krypton and xenon detectors, which are bright (UV) scintillators \cite{Ada13}. Other homogeneous detectors of electromagnetic (em) showers are based on detection of \v{C}erenkov light, in particular 
lead glass \cite{Akr90}. Very large water \v{C}erenkov calorimeters (\eg SuperKamiokande \cite{Fuk03}) should also be mentioned in this category.

Sampling calorimeters, which are typically much cheaper, become competitive at higher energies.
In properly designed instruments of this type, the energy resolution is usually dominated by {\sl sampling fluctuations}. 
In that context, an example of a non-properly designed instrument is a very-fine-sampling $10 X_0$ deep calorimeter intended for detecting 1 TeV photons. The energy resolution of such a calorimeter would be dominated by fluctuations in the energy fraction leaking out from the back, and the contribution of sampling fluctuations would be marginal at best.

Sampling fluctuations represent fluctuations in the number of different shower particles that contribute to the  calorimeter signals, convoluted with fluctuations in the amount of energy deposited by individual shower particles in the active calorimeter layers. They depend both on the {\sl sampling fraction}, which is determined by  the ratio of active and passive material, and on the {\sl sampling frequency}, determined by the number of different sampling elements in the region where the showers develop. 
Sampling fluctuations are stochastic and their contribution to the relative energy resolution, $\sigma/E$, is described by \cite{Liv95}
\vskip 1mm
\begin{equation}
{\sigma \over E}~= {a \over \sqrt{E}}~~~~~~{\rm with} ~~~~~a~=~0.027 \sqrt{d/f_{\rm samp}}
\label{eq:reso}
\end{equation}
in which $d$ represents the thickness (or diameter, in the case of fibers) of individual active sampling layers (in mm), $f_{\rm samp}$ the sampling fraction for minimum ionizing particles ({\sl mip}s), and $E$ is the particle energy in GeV.
This expression describes data obtained with a large variety of different (non-gaseous) sampling calorimeters reasonably well.

Another factor that may contribute to the em energy resolution of a sampling calorimeter is determined by the number of signal quanta that constitutes the signal. Fluctuations in that number are usually negligible when the signal consists of electrons produced in ionization processes (\eg in calorimeters using liquid argon, krypton or xenon as active medium), or when scintillation light is the origin of the signals. In these cases, the number of signal quanta amounts typically to at least several hundred per GeV deposited energy, so that the contribution of fluctuations in that number to the energy resolution amounts to $\lesssim 5\%/\sqrt{E}$, with $E$ expressed in GeV. Such fluctuations are also stochastic and their contribution to the energy resolution thus has to be combined in quadrature with that from sampling fluctuations.

Fluctuations in the number of signal quanta may dominate the em energy resolution of a sampling calorimeter when \v{C}erenkov light is the origin of the signals. For example, in the dual-readout fiber calorimeters discussed in Sections VI and VII, the \v{C}erenkov light yield is of the order of 30 photoelectrons per GeV deposited energy, which translates into fluctuations of the order of $17\%/\sqrt{E}$,
This is a non-negligible contribution to the total em energy resolution of these instruments, in which the signals from the scintillation and \v{C}erenkov fibers are combined.

\subsection{Hadron calorimeters}

The energy range covered by hadron calorimeters is in principle even larger than that for em ones. Calorimetric techniques are
used to detect thermal neutrons, which have kinetic energies of a small fraction of 1 eV, to the highest-energy particles
observed in nature, which reach the Earth from outer space as cosmic rays carrying up to $10^{20}$ eV or more. In accelerator-based particle physics
experiments, hadron calorimeters are typically used to detect protons, pions, kaons and fragmenting quarks and gluons (commonly referred to as {\sl jets}) with energies in the GeV - TeV range. In this paper, we mainly discuss the latter instruments.

The development of hadronic cascades in dense matter differs in essential ways from that of electromagnetic ones, with important consequences for calorimetry.
Hadron showers consist of two distinctly different components:
\begin{enumerate}
\item An {\sl electromagnetic} component; $\pi^0$ and $\eta$ mesons generated in the absorption process
decay into photons which develop em showers.
\item A {\sl non-electromagnetic} component, which comprises essentially everything else that
takes place in the absorption process. 
\end{enumerate}
For the purpose of calorimetry, the main difference between these components is that
some fraction of the energy contained in the non-em component does {\sl not} contribute to the signals. This {\em invisible energy}, which mainly consists of the binding energy of nucleons released in the numerous 
nuclear reactions, may represent up to 40\% of the total non-em energy, with large event-to-event fluctuations.

The appropriate length scale of hadronic showers is the nuclear interaction length ($\lambda_{\rm int}$), 
which is typically much larger (up to 30 times for high-$Z$ materials) than the radiation length ($X_0$), which governs the development of em showers.
Many experiments make use of this fact to distinguish between electrons and hadrons on the basis of the energy deposit profile
in their calorimeter system. Since the ratio $\lambda_{\rm int}/X_0$ is proportional to $Z$, particle identification on this basis works best for high-$Z$ absorber materials. Lead, tungsten and depleted uranium are therefore popular choices for the absorber material in preshower detectors and the first section of a longitudinally segmented calorimeter,
which is therefore commonly referred to as the {\sl electromagnetic section}.

Just as for the detection of  em showers, high-resolution hadron calorimetry requires an average longitudinal containment better than 99\%. 
In iron and materials with similar $Z$, which are most frequently used for hadron calorimeters, 99\% longitudinal containment requires a thickness ranging from $5 \lambda_ {\rm int}$ for a particle energy of 20 GeV to $8 \lambda_{\rm int}$ at 150 GeV.  Hadronic energy resolutions of the order of 1\% require not only longitudinal shower containment at the 99\% level, but also lateral containment of 98\% or better.

Energetic $\pi^0$s may be produced throughout the absorber volume, and not exclusively in the em calorimeter section. They lead to local regions of highly concentrated energy deposit. Therefore,
there is no such thing as a ``typical hadronic shower profile". This feature affects not only the shower containment requirements, but also the
calibration of longitudinally segmented calorimeters in which one tries to improve the quality of calorimetric energy
measurements of jets with an upstream tracker, which can measure the momenta of the charged jet constituents with great precision, \ie the {\sl Particle Flow Analysis} method mentioned in Section I.

\subsection{{Compensation}}

\begin{figure}[hbtp]
\epsfysize=10.6cm
\centerline{\epsffile{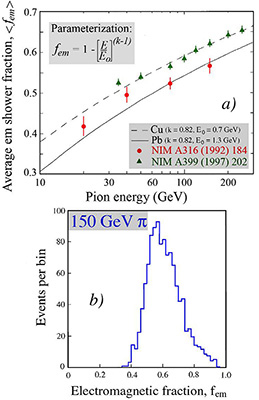}}
\caption{\small
The energy dependence of the average em shower fraction measured for copper- and lead-based sampling calorimeters ($a$), and the event-to-event fluctuations in this fraction, for showers induced by 150 GeV $\pi^-$ in lead ($b$). Experimental data from \cite{Aco92a} (lead) and \cite{Akc97} (copper).}  
\label{femprops}
\end{figure}

The properties of the em shower component have also important consequences for the {\em energy resolution}, the signal {\em linearity} and the {\em response function}.
The average fraction of the total shower energy contained in the em component has been measured to increase
with energy following a power law:
\begin{equation}
\langle f_{\rm em} \rangle ~=~1 - \bigl[E/E_0\bigr]^{k-1}
\label{femE}
\end{equation}
where $E_0$ is a material dependent constant related to the average multiplicity in hadronic interactions (varying from 0.7 GeV to 1.3 GeV for $\pi$-induced reactions on Cu and Pb, respectively), and $k \sim 0.82$ \cite{Gab94}. This is illustrated in Figure \ref{femprops}a. For proton-induced reactions, $\langle f_{\rm em} \rangle$ is typically considerably smaller than for pion-induced ones, as a result of baryon number conservation in the shower development, which prevents the production of a leading $\pi^0$ \cite{Akc98}.

Let us define the calorimeter {\em response} as the conversion efficiency from deposited energy to generated signal, and normalize it to that for minimum ionizing particles. The calorimeter response to showers is usually different from that to {\sl mip}s.
\begin{figure}[hbtp]
\epsfxsize=8cm
\centerline{\epsffile{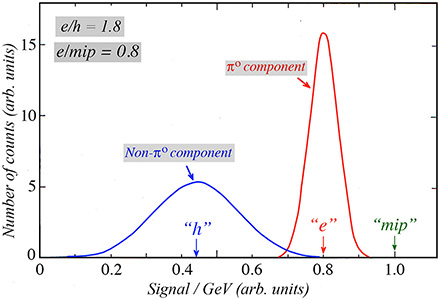}}
\caption{\small
Illustration of the meaning of the $e/h$ and $e/mip$ values of a (generic) calorimeter. Shown are distributions of the signal per unit deposited energy for the electromagnetic and non-em components of hadron showers. These distributions are normalized to the response for minimum ionizing particles ($``mip"$). The average values of the em and non-em distributions are the em response ($``e"$) and non-em response ($``h"$) , respectively.}  
\label{ehprinciple}
\end{figure}
The responses of a given calorimeter to the em and non-em hadronic shower components, $e$ and $h$, are usually not the same
either, as a result of invisible energy and a variety of other effects. Such calorimeters are called {\em non-compensating} ($e/h \ne 1$).
Since their response to hadrons, $\Bigl[\langle f_{\rm em} \rangle + \bigl[1 -  \langle f_{\rm em} \rangle\bigr] h/e\Bigr] (e/mip)$, is energy dependent (\ref{femE}), they are intrinsically non-linear. The meaning of the different aspects of the calorimeter response is illustrated in Figure \ref{ehprinciple}. The calorimeter is characterized by the $e/h$ and $e/mip$ ratios, which in this example have values of 1.8 and 0.8, respectively.

Event-to-event fluctuations in $f_{\rm em}$ are large and non-Poissonian (Figure \ref{femprops}b). If $e/h \ne 1$, these fluctuations 
tend to dominate the hadronic energy resolution and their asymmetric distribution characteristics are reflected in the 
response function \cite{Liv17}.

The effects of non-compensation on the hadronic energy resolution, linearity and line shape, as well as the associated calibration problems \cite{Olga}, are absent in compensating calorimeters ($e/h = 1.0$). Compensation can be achieved in sampling calorimeters with high-$Z$ absorber material and hydrogenous active material. It requires a very specific sampling fraction, so that the response to shower neutrons is boosted by the precise factor needed to equalize $e$ and $h$. For example, in Pb/scintillating-plastic structures, this sampling fraction is $\sim 2\%$ for showers \cite{Bern87,Aco91c,Suz99}. This small
sampling fraction sets a lower limit on the contribution of sampling fluctuations, while the need to detect MeV-type neutrons with high efficiency requires signal integration over a relatively large volume during at least 30 ns. Yet, calorimeters of this type currently hold the world record for hadronic energy resolution \cite{Aco91c}.

\subsection{Calorimetry in the TeV regime}

An often mentioned design criterion for calorimeters at a future high-energy linear $e^+e^-$ collider is the need to distinguish between hadronically decaying 
$W$ and $Z$ bosons \footnote{An
important reaction to be studied is $e^+e^- \rightarrow H^0Z^0$. By using the hadronic decay modes of the $Z^0$ (in addition to $e^+e^-$ and $\mu^+\mu^-$ decay), an important gain in event rates can be obtained. However, more abundant processes such as $e^+e^- \rightarrow W^+W^-$ will obscure the signal unless the calorimeter is able to distinguish efficiently between hadronic decays of $W$ and $Z$ bosons.}.  The requirement that the di-jet masses of $W \rightarrow q\bar{q}^\prime$ and $Z \rightarrow q\bar{q}$ events are separable by at least one Rayleigh criterion
implies that one should be able to detect hadronic energy deposits of 80 - 90 GeV with a resolution of 3 - 3.5 GeV.
This goal can be, and has been achieved with compensating calorimeters for single hadrons \cite{Aco91c,Beh90}, but not for jets.
However, because of the small sampling fraction required for compensation, the em energy resolution is somewhat limited in such devices. And because of the crucial role of neutrons produced in the shower development, the signals would have to be integrated over relatively large volumes and time intervals to achieve this resolution, which is not always possible in practice.
In this paper, we discuss a new technique that is currently being pursued to circumvent these limitations. However, first, we briefly review the factors that determine and limit the hadronic calorimeter resolution.
\vskip 2mm

In the TeV domain, it is incorrect to express calorimetric energy resolutions in terms of $a/\sqrt{E}$, as is often done. Deviations from $E^{-1/2}$ scaling are the result of non-Poissonian fluctuations. These manifest themselves typically predominantly at high energies, where the contribution of the Poissonian component becomes very small. 
It is often assumed that the effect of non-compensation on the energy resolution is energy independent (``constant term''). 
It turns out that the effects of 
fluctuations in the em shower fraction, $f_{\rm em}$, are more correctly described by a term that is very similar to the one used to describe the energy dependence of $\langle f_{\rm em}\rangle$. This term should be added in quadrature to the $E^{-1/2}$ scaling term which accounts for the Poisson fluctuations \cite{PDG16}:

\begin{equation}
{\sigma\over E}~=~{a_1\over \sqrt{E}} \oplus a_2 \biggl[ \biggl({E\over E_0}\biggr)^{l-1}\biggr]
\label{lognoncomp2}
\end{equation}
Just as in Equation \ref{femE}, $E_0$ is a material dependent constant related to the average multiplicity in the hadronic interactions
and $l$ (which has a value of 0.72 in copper) is determined by the energy dependence of $\langle f_{\rm em}\rangle$ \cite{Gab94}. The parameter $a_2$ is determined by the degree of non-compensation. It varies between 0 (for compensating calorimeters) and 1 (for extremely non-compensating calorimeters). Following Groom \cite{Gro07}, we assume a linear relationship for intermediate $e/h$ values:

\begin{equation}
a_2 = |1 - h/e|
\end{equation}
\begin{figure}[htb]
\epsfysize=11.9cm
\centerline{\epsffile{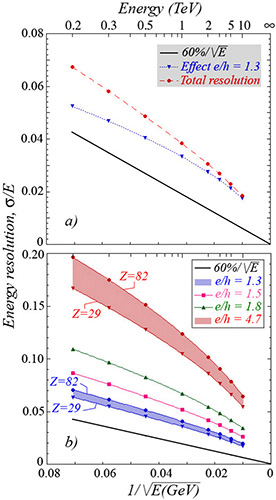}}
\caption{\small Hadronic energy resolution in the TeV domain, calculated with Equation \ref{lognoncomp2} \cite{Wig08}.} 
\label{ehres}
\end{figure}
Experiments with several calorimeters (\eg \cite{Ari94}) have revealed that the hadronic energy resolution data are well described by a {\sl linear sum} of a stochastic and constant term: 

\begin{equation}
{\sigma\over E}~=~{c_1 \over \sqrt{E}} + c_2
\label{linres}
\end{equation}
which is, in the energy range covered by the current generation of test beams, \ie up to 400 GeV, not distinguishable from Equation \ref{lognoncomp2}, albeit that the stochastic parameters differ ($c_1 > a_1$).
Interestingly, the similarity between the two expressions disappears when the energy range is extended into the TeV domain. This is illustrated in Figure \ref{ehres} \cite{Wig08}. The curves in these graphs represent Equation \ref{lognoncomp2} for the energy range from 0.2 - 10 TeV. Figure \ref{ehres}a shows the contributions of the stochastic and the non-compensation term as a function of energy, as well as the total energy resolution, for a calorimeter with $e/h = 1.3$ and a stochastic term of $60\%/\sqrt{E}$. It is clear that, even for an $e/h$ value that is usually considered quite good, the effects of fluctuations in $f_{\rm em}$ dominate the hadronic energy resolution
in the TeV regime. Figure \ref{ehres}b shows the total energy resolution for calorimeters with different $e/h$ values. Especially for large $e/h$ values, the energy dependence of the resolution is no longer well described
by a straight line in this plot (thus invalidating Equation \ref{linres}). Figure \ref{ehres}b also shows the effects that may be expected as a result of material dependence. These derive from the value of $E_0$ in Equation \ref{lognoncomp2}, which is almost a factor of two larger in high-$Z$ absorber materials such as lead, compared to copper or iron.

\section{The principles of dual-readout calorimetry}

A major intrinsic problem for the performance of hadron calorimeters is the fact that nuclear binding energy losses, and event-to-event fluctuations in this {\sl invisible energy}, lead to 
differences in the response functions to the em and non-em shower components ($e/h \ne 1$, see Figure \ref{ehprinciple}) \footnote{Especially in experiments at hadron colliders, other factors may cause major problems as well, for example event pileup. In practice, the experiments at the Tevatron and the LHC have found ways to mitigate the consequences of effects that limit the hadronic calorimeter performance, for example by incorporating the experimental information from other detector components.}. 
In compensating calorimeters, the responses to the em and non-em components are equalized by design, which may lead to a substantial improvement of the performance. 

\begin{figure}[htb]
\epsfxsize=8cm
\centerline{\epsffile{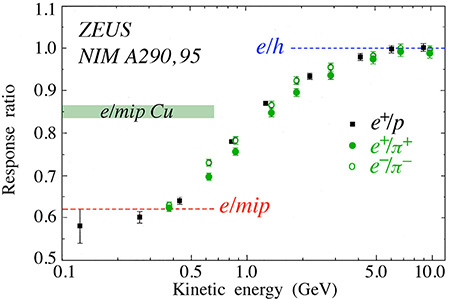}}
\caption{\small The response of the uranium-based ZEUS calorimeter to low-energy hadrons. Experimental data points from \cite{And90}. For comparison, the e/mip ratio for a copper-based calorimeter is shown as well.} 
\label{zeuslin2}
\end{figure}

\begin{figure}[htbp]
\epsfysize=11cm
\centerline{\epsffile{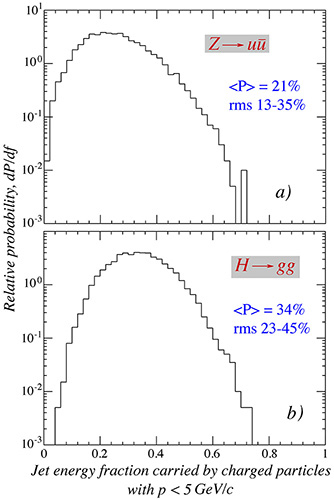}}
\caption{\small Distribution of the fraction of the energy released by hadronically decaying $Z^0$ ($a$) and $H^0$ ($b$) bosons
at rest that is carried by charged final-state particles with a momentum less than 5 GeV/$c$ \cite{Web15}. See text for details.} 
\label{webber}
\end{figure}

As stated above, the main drawbacks of compensating calorimeters derive from the need for a high-$Z$ absorber material, such as lead or uranium. This absorber material both reduces the em response and generates a large number of neutrons, the two ingredients that are crucial for achieving the compensation condition, $e/h = 1.0$, and thus for eliminating the contribution of fluctuations in the em shower fraction, $f_{\rm em}$. However, the small $e/mip$ value, typically $\sim 0.6$ in these absorber materials, leads to large response non-linearities for low-energy hadrons. These particles lose their kinetic energy predominantly through ionization of the absorber medium, rather than through shower development (Figure \ref{zeuslin2}). 
Such particles account for a significant fraction of the energy of high-energy jets, such as the ones produced in the hadronic decay of the $W$ and $Z$ intermediate vector bosons. Figure \ref{webber} shows the distribution of the energy released by $Z^0$s (decaying through the process $Z^0 \rightarrow u\bar{u}$) and Higgs bosons (decaying into a pair of gluons) at rest, that is carried by charged final-state particles with a momentum less than 5 GeV/$c$. 
The figure shows that, on average, 21\% of the energy equivalence of the $Z^0$ mass is carried by such particles. The event-to-event fluctuations are such that this fraction varies between 13\% and 35\% in a $1 \sigma_{\rm rms}$ interval around this mean value, \ie for 68.27\% of the $Z$ decays, the fraction of the $Z$ mass energy carried by these soft fragments is somewhere between 13\% and 35\% . For Higgs bosons decaying into a pair of gluons, the average fraction is even larger, 34\%, with rms variations between 23\% and 45\%.
A gluon jet has a higher average multiplicity than a jet resulting from a fragmenting quark or anti-quark, because of differences between the color factors in the parton branching and differences in the fragmentation function, \eg the absence of a leading-particle effect \cite{Web15}. 

As a result of the important contribution from soft jet fragments, and the large event-by-event fluctuations in this contribution, the energy resolution for intermediate vector bosons measured with the compensating ZEUS uranium calorimeter was worse than expected on the basis of the single-pion resolution. 
Also, the small sampling fraction required to achieve compensation limited the em energy resolution, to $18\%/\sqrt{E}$ in ZEUS. And the (properly amplified) contributions of neutrons, which are equally essential for this purpose, made it necessary to integrate the hadronic signals over a rather large time interval ($\gtrsim 30$ ns) and calorimeter volume ($\sim 1$ m$^3$).

An alternative approach to eliminate the effects of the fluctuations in the em shower fraction, which dominate the hadronic energy resolution of non-compensating calorimeters, is to {\sl measure} $f_{\rm em}$ for each event. It turns out that the \v{C}erenkov mechanism provides unique opportunities to achieve this.

Calorimeters that use \v{C}erenkov light as signal source are, for all practical purposes, only
responding to the em fraction of hadronic showers \cite{Akc97}. This is because the electrons/positrons through which the energy is deposited in the em shower component are relativistic down to energies
of only $\sim$200 keV. On the other hand, most of the non-em energy in hadron showers is deposited by
non-relativistic protons generated in nuclear reactions.
Such protons do generate signals in active media such as plastic scintillators or liquid argon. By comparing the relative strengths of the 
signals representing the visible deposited energy and the \v{C}erenkov light produced in the shower absorption process, the em shower fraction can be determined and the total shower energy can be reconstructed using the known $e/h$ value(s) of the calorimeter. This is the essence of what has become known as {\sl dual-readout }calorimetry.

The Dual-REAdout Method (DREAM) allows the elimination of the
mentioned drawbacks of intrinsically compensating calorimeters:
\begin{enumerate}
\item There is no reason to use high-$Z$ absorber material. An absorber such as copper has an $e/mip$ value of 0.85, which strongly mitigates the effects of non-showering hadrons on the jet energy resolution. Also, by using copper instead of lead or uranium,  a calorimeter 
with a given depth (expressed in nuclear interaction lengths) will need to be much less massive.
\item The sampling fraction of detectors based on this method can be chosen as desired. As a result, excellent em
energy resolution is by no means precluded.
\item The method does not rely on detecting neutrons (although these may offer some additional advantages, as shown in the following).
Therefore, there is no need to integrate the signals over large times and detector volumes.
\end{enumerate}

\section{The initial attempt: ACCESS}   

The idea to use the complementary information from scintillation and \v{C}erenkov light was first applied in
a prototype study for ACCESS, a high-energy cosmic-ray experiment proposed for the International Space 
Station \cite{Nag01} \footnote{The ACCESS project was canceled after the 2003 accident with the Columbia space shuttle.}. 
Because of the very severe
restrictions on the mass of the instruments, the ACCESS calorimeter had to be very thin, less than 2
$\lambda_{\rm int}$. It was therefore imperative to maximize the amount of information obtained per unit
detector mass.
\begin{figure*}[hbt]
\epsfysize=6.4cm
\centerline{\epsffile{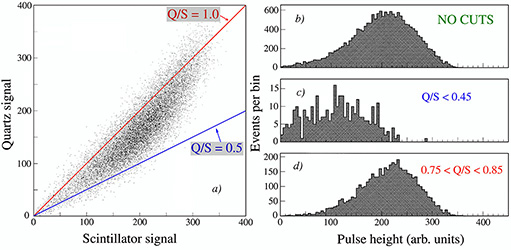}}
\caption{\small
Results of tests of the dual-readout ACCESS calorimeter with 375 GeV pions. Scatter plot of the
signals recorded in the quartz fibers \vs those in the scintillating fibers ($a$). The signal
distributions from the scintillating fibers for all events ($b$) and for subsets of events with
a small ($c$) or average ($d$) fraction of \v{C}erenkov light \cite{Nag01}.}
\label{access2}
\end{figure*}

When high-energy hadrons develop showers in such a thin calorimeter, the response function is
completely determined by leakage fluctuations. These fluctuations are very likely correlated with the
fraction of energy spent on $\pi^0$ production inside the detector. In general, $\pi^0$s
produced in the first nuclear interaction develop em showers that are contained in the
detector, while charged pions typically escape. Therefore, events in which a large fraction of
the initial energy is converted into $\pi^0$s in the first interaction will exhibit little
leakage (\ie a large detector signal), while events in which a small fraction of the energy has
been transferred to $\pi^0$s will be characterized by large leakage (\ie small detector signals).
A dual-readout calorimeter that measures both the ionization losses ($dE/dx$) and the production of
\v{C}erenkov light might distinguish between events with relatively small and large shower
leakage, since the ratio of the two signals would be different in these two cases: A relatively large
\v{C}erenkov signal would indicate relatively little shower leakage, while a small \v{C}erenkov signal
(compared to the $dE/dx$ signal) would suggest that a large fraction of the shower energy escaped from
the detector.

The dual-readout calorimeter prototype built for ACCESS consisted of a 1.4 $\lambda_{\rm int}$ deep lead absorber structure, in which alternating ribbons of two types of optical fibers were embedded. The signals from the scintillating fibers provided a measure for the total energy deposited by the showers, while quartz fibers recorded the \v{C}erenkov light produced in the absorption process.
Figure \ref{access2} shows some results of the tests of this instrument. These tests were carried out at CERN with a beam of 375 GeV pions. In Figure
\ref{access2}a,  the signals recorded by the quartz fibers are plotted versus those from the scintillating
fibers. The non-linear correlation between these signals indicates that they
indeed measured different characteristics of the showers.

The scintillation signal distribution, \ie the projection of the scatter plot on the horizontal axis,
is shown in Figure \ref{access2}b. The fact that this distribution is skewed to the low-energy side may
be expected as a result of shower leakage. The {\em ratio of the signals from the quartz fibers
and from the scintillating fibers ($Q/S$)} corresponds to the slope of a line through the bottom left corner
of Figure \ref{access2}a. The two lines drawn in this figure represent $Q/S =1$ and $Q/S = 0.5$, respectively.

In Figure \ref{access2}c, the signal distribution is given for events with a small $Q/S$ value ($Q/S <
0.45$). These events indeed populate the left-side tail of the calorimeter's response function (Figure
\ref{access2}b). This distribution is very different from the one obtained for events with $Q/S$ ratios
near the most probable value, shown in Figure \ref{access2}d. 
The average values of the scintillation signal distributions in Figures \ref{access2}c and
\ref{access2}d differ by about a factor of two.

These results demonstrate that events from the tails of the 
$Q/S$ distribution correspond to events from the tails of the ($dE/dx$) response function.
Therefore, the ratio of the signals from the quartz and the scintillating
fibers does indeed provide information on the energy containment and may thus be used to reduce the
fluctuations that dominate the response function of this very thin calorimeter.
The authors showed that the resolution could be improved by $\sim$10-15\% using the $Q/S$ information and that this
improvement was primarily limited by the small light yield of the quartz fibers, 0.5 photoelectrons per GeV.
Fluctuations in the number of
\v{C}erenkov photoelectrons determined the width of the ``banana'' in Figure \ref{access2}a and thus the
selectivity of $Q/S$ cuts. Therefore, the relative improvement in the energy resolution also increased with the
hadron energy. 

It is remarkable that the dual-readout technique already worked so well in this very thin calorimeter. After all, in this
detector one is looking only at the very first generation of shower particles and the non-em shower
component has barely had a chance to develop. The overwhelming majority of the non-relativistic shower
particles, in particular the spallation and recoil protons, are produced in later stages of the hadronic shower
development. The signals from these non-relativistic shower particles are crucial for the success of the
method, since they are the ones that do produce scintillation light and no \v{C}erenkov light.       
The fact that the technique already appeared to work so well in this very thin calorimeter therefore held the
promise that excellent results might be expected for detectors that fully contain the showers. 

\section{Dual-readout data analysis procedures}

A dual-readout calorimeter produces two types of signals for the showers developing in it, a scintillation signal ($S$) and a \v{C}erenkov signal ($C$). Both signals can be calibrated with electrons of known energy $E$, so that $\langle S\rangle = \langle C\rangle = E$ for em showers,
and the calorimeter response to em showers, $R_{\rm em} = \langle S\rangle /E = \langle C\rangle /E = 1$.
For a given event, the hadronic signals of this calorimeter can then be written as
\begin{eqnarray}
S~=~ E\Bigl[f_{\rm em} + {1\over {(e/h)_S}} (1 - f_{\rm em})\Bigr] \nonumber \\
C~=~ E\Bigl[f_{\rm em} + {1\over {(e/h)_C}} (1 - f_{\rm em})\Bigr]
\label{eq2}
\end{eqnarray}
%
%
\ie as the sum of an em shower component ($ f_{\rm em}$) and a non-em shower component (1 -- $f_{\rm em}$). The contribution of the latter component to the reconstructed energy is weighted by a factor $h/e$. When $f_{\rm em} = 1$ or $e/h = 1$, the hadronic shower response is thus the same as for electrons: $R = 1$. However, in general $f_{\rm em} < 1$ and $e/h \ne 1$, and therefore the hadronic response is different from 1. The reconstructed energy is thus different (typically smaller) than $E$.

The dual-readout method works thanks to the fact that $(e/h)_S \ne (e/h)_C$. The larger the difference between both values, the 
better. The em shower fraction $f_{\rm em}$ and the shower energy $E$ can be found by solving Equations \ref{eq2}, using the measured values of the scintillation and \v{C}erenkov signals and the {\bf known} $e/h$ ratios of the \v{C}erenkov and scintillator calorimeter structures. We will describe later {\sl how} these ratios can be determined. 

\begin{figure}[htbp]
\epsfxsize=8cm
\centerline{\epsffile{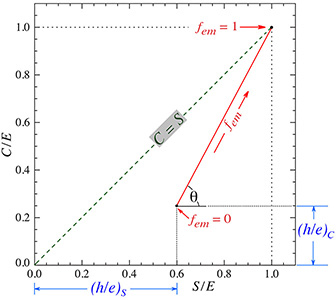}}
\caption{\small
Graphic representation of Equations \ref{eq3} \cite{PDG16,LLW17}. The data points for hadron showers detected with a dual-readout calorimeter are located around the straight (red) line in this diagram. The data points for em showers in this calorimeter are clustered around the point where this line intersects the $C = S$ line, \ie the point (1,1). See text for further details.}
\label{dr1}
\end{figure}
Looking at Equations \ref{eq2}, we see that the ratio of the two measured signals $S$ and $C$ is {\sl independent of the shower energy} $E$.
There is thus a one-to-one correspondence between this measured signal ratio and the value of the em shower fraction, $f_{\rm em}$.
This fraction can thus be determined for each individual event, and therefore the effects of fluctuations in $f_{\rm em}$ can be eliminated.
Just as in compensating calorimeters, where these fluctuations are eliminated by design, this is the most essential ingredient for improving the quality of hadron calorimetry.

\begin{figure*}[htbp]
\epsfysize=8.6cm
\centerline{\epsffile{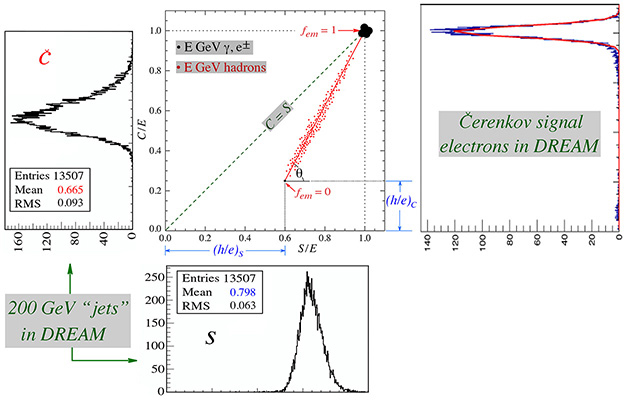}}
\caption{\small
The $S-C$ diagram of the signals from a (generic) dual-readout calorimeter \cite{LLW17}. The hadron events are clustered around the straight (red) line, the electron events around the point (1,1).
Experimental signal distributions measured in the scintillation and \v{C}erenkov channels for 200 GeV ``jets''
with the DREAM fiber calorimeter \cite{Akc05b} are shown as well. Also shown is a typical (\v{C}erenkov) response function measured for electrons in DREAM.}
\label{dr2}
\end{figure*}

Let us now look again at Equations \ref{eq2}, and rewrite these as
\begin{eqnarray}
S/E~=~ (h/e)_S + f_{\rm em}\bigl[1 - (h/e)_S\bigr] \nonumber \\
C/E~=~ (h/e)_C + f_{\rm em}\bigl[1 - (h/e)_C\bigr]
\label{eq3}
\end{eqnarray}
Figure \ref{dr1} shows that the experimental data points for hadron showers detected with a dual-readout calorimeter are thus located around a straight (red) line in the $C/E$ \vs $S/E$ diagram. This line links the points $[(h/e)_S,(h/e)_C]$, for which $f_{\rm em} = 0$, with the point
(1,1), for which $f_{\rm em} = 1$. The experimental data points for electron showers are concentrated around the latter point, as illustrated in Figure \ref{dr2}.

The $f_{\rm em}$ value for an individual hadron event is directly related to the ratio of the two signals ($C/S$) and can be found by solving Equations \ref{eq3}, using the known values of $(h/e)_S$ and $(h/e)_C$:
\begin{equation}
f_{\rm em} = {{(h/e)_C - (C/S) (h/e)_S}\over {(C/S) [1 - (h/e)_S] - [1 - (h/e)_C] }} 
\label{eq3a}
\end{equation}

Figure \ref{dr2} shows, apart from the $S-C$ diagram, also some signal distributions obtained with the dual-readout calorimeter described in Section VI.
The scintillation and \v{C}erenkov signal distributions measured for 200 GeV multiparticle events (to be called ``jets" in the following) \footnote{The calorimeter performance for these objects was studied with events created by means of interactions of beam particles in a target placed upstream of the calorimeter. Typically, these events were required to have a certain minimum multiplicity. These {\sl multiparticle} events are, of course, not the same as the QCD jets that originate from a fragmenting quark or gluon. Yet, for the purpose of calorimetry they are very useful, since they represent a collection of particles that enter the calorimeter simultaneously. The composition of this collection is unknown, but the total energy is known. In the absence of a jet test beam, this is a reasonable alternative.} are the projections of the data points ({\sl S,\v{C}}) on the horizontal and vertical axes of the diagram, respectively. Their asymmetric shape reflects the asymmetric $f_{\rm em}$ distributions (see Figure \ref{femprops}b). The electron showers measured 
with this detector, both in the scintillation and the \v{C}erenkov channels, are centered around the point (1,1) in this plot.

\begin{figure*}[htbp]
\epsfysize=9.1cm
\centerline{\epsffile{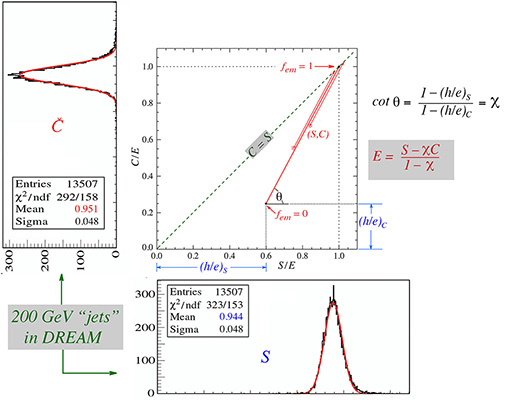}}
\caption{\small
The $S-C$ diagram of the signals from a (generic) dual-readout calorimeter \cite{LLW17}. The hadron events are clustered around the straight (red) line, the electron events around the point (1,1).
Experimental signal distributions measured in the scintillation and \v{C}erenkov channels for 200 GeV ``jets'' with the DREAM fiber calorimeter, after applying the dual-readout transformation (\ref{eq5}) are shown as well \cite{Akc05b}. }
\label{dr3}
\end{figure*}

The slope of the red line around which the hadron data points are clustered, \ie the angle $\theta$, only depends of the two $e/h$ values, and is thus {\sl independent of the hadron energy}. We define 

\begin{equation}
\cot \theta = {{1 - (h/e)_S}\over {1 - (h/e)_C}} = \chi
\label{eq4}
\end{equation}
and the parameter $\chi$ is thus also independent of energy. Because of this feature, the scintillation and \v{C}erenkov signals measured for 
a particular hadron shower can be used to reconstruct its energy in an unambiguous way:

\begin{equation}
E~=~ {{S - \chi C}\over {1 - \chi}}
\label{eq5}
\end{equation}
This is graphically illustrated in Figure \ref{dr3}, since Equation \ref{eq5} implies that the data point ($S,C$) is moved up along the red straight line 
until it intersects the line defined by $C = S$. If this is done for all hadronic data points, the result is a collection of data points that cluster around the point (1,1), just like the data points for electron showers. 

The effect of this operation on the experimental signal distributions from Figure \ref{dr2} is also displayed in Figure \ref{dr3}, which shows that these distributions have become 
much more narrow, well described by Gaussian functions and centered close to the same value as em showers (0.951, 0.944 \vs 1). The 5\% difference in the reconstructed energy is in this case most likely due to the fact that these data concern multiparticle events produced by interactions in a target upstream of the calorimeter.
 
The dual-readout procedure thus effectively uses the measured signals to determine the em shower fraction, $f_{\rm em}$, and then calculates what the signals would be if $f_{\rm em}$ was 1.0. The actual $f_{\rm em}$ distribution for showers produced in the absorption of a sample of hadrons of the same type and energy is therefore not a factor that affects the energy measurement for that event sample. A dual-readout calorimeter is therefore {\sl linear} for hadron detection, since the correct energy is reproduced in each case. 
\begin{figure}[b!]
\epsfxsize=8cm
\centerline{\epsffile{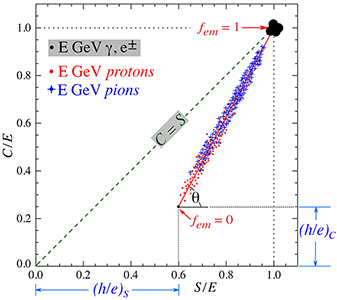}}
\caption{\small
The $S-C$ diagram of the signals from a (generic) dual-readout calorimeter. The hadron events are clustered around the straight (red) line. Data points for protons and pions have different distributions, reflecting differences in the em shower fraction \cite{LLW17}.}
\label{dr4}
\end{figure}

Interestingly, a dual-readout calorimeter will also produce signal distributions with the same average value for event samples of pions, protons and kaons of the same energy. The $f_{\rm em}$ distributions are quite different for showers produced by these different types of hadrons, as a result of conservation of baryon number and strangeness in the shower development. This prevents the production of a very energetic, leading $\pi^0$ in the case of protons and kaons, respectively. Measurements with conventional calorimeters have clearly shown significant differences between the response functions of protons and pions. Response differences of $\sim 5\%$ have been reported by ATLAS \cite{Adr09}, while differences in the CMS Forward Calorimeter exceeded 10\% for energies below 100 GeV \cite{Akc98}. This feature translates into a systematic uncertainty in the hadronic energy measurement, unless one knows what type of hadron caused the shower (which at high energies is, in practice, rarely the case). 
Figure \ref{dr4} illustrates that the mentioned effects do not play a role for dual-readout calorimeters. The relationship \ref{eq5} is universally valid for all types of hadrons, and also for jets.

The fact that $\theta$ and $\chi$ are independent of the energy and the particle type offers an interesting possibility to measure the hadronic energy with unprecedented precision, at least for an ensemble of particles with the same energy. In practice, the energy resolution is usually determined in that way, \ie as the fractional width ($\sigma/E$) of the signal distribution for a beam of mono-energetic particles produced by an accelerator.

\begin{figure}[htb]
\epsfxsize=7.5cm
\centerline{\epsffile{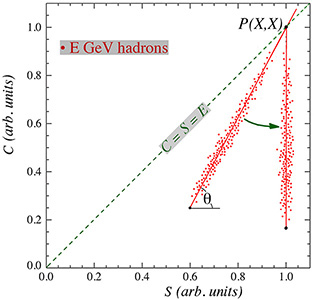}}
\caption{\small
The $S-C$ diagram of the signals from a (generic) dual-readout calorimeter. The hadron events are clustered around the straight (red) line. Also shown is the effect of a rotation of this red line and the associated distribution of data points \cite{LLW17}. }
\label{dr5}
\end{figure}

The so-called ``rotation method''  \cite{LLW17} works as follows (see Figure \ref{dr5}). First, the experimental hadronic data points are fitted with a straight line. This line intersects the $C = S$ line at point $P (X,X)$. Since this point represents hadron showers for which $f_{\rm em} =1$, data points for electrons with the same energy as the hadrons are in principle clustered around the same point in the $S-C$ diagram.
Next, the measured distribution of the hadronic data points is rotated around point $P$ (to which the coordinates (0,0) are assigned for this purpose), over an angle $90^\circ - \theta$. This procedure corresponds to a coordinate transformation of the type

\begin{equation}
\begin{pmatrix} S^\prime \\ C^\prime \end{pmatrix} \quad
=~~
\begin{pmatrix} \sin{\theta}&-\cos{\theta} \\ \cos{\theta}&\sin{\theta} \end{pmatrix} \quad
\begin{pmatrix} S \\ C \end{pmatrix} \quad \\
\label{rot}
\end{equation}
After accounting for the frame translation, the new coordinates of the data points thus become $(S^\prime + X, C^\prime +X)$, where $X$ is derived from the fit
of the $(S,C)$ data points. 
The projection of the rotated scatter plot on the $x$-axis is a narrow signal distribution centered around the (approximately) correct energy value.  
\begin{figure}[htbp]
\epsfysize=17cm
\centerline{\epsffile{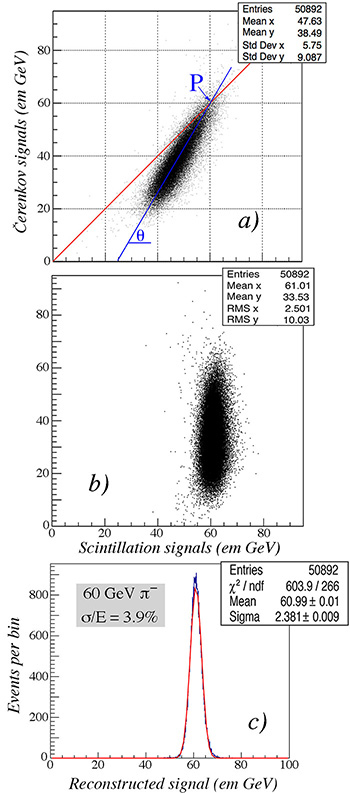}}
\caption{\small Signal distributions of the RD52 Dual-Readout lead-fiber calorimeter for 60 GeV pions. Scatter plot of the two types of signals as recorded for these particles ($a$) and rotated over an angle $\theta = 30^\circ$ around the point where the two lines from diagram $a$ intersect ($b$). Projection of the latter scatter plot on the $x$-axis ($c$). Data from \cite{Lee17}.}
\label{DRrotation}
\end{figure}

Figure \ref{DRrotation} shows an example of results obtained in practice with a procedure of this type \cite{Lee17}, for a beam of 60 GeV $\pi^-$.
This resulting signal distribution is well described by a Gaussian function with a central value of 61.0 GeV and a relative width, $\sigma/E$, of $3.9\%$. This corresponds to $30\%/\sqrt{E}$. The narrowness of this distribution reflects the clustering of the data points around the axis of the locus in Figure \ref{DRrotation}a. It should be pointed out that the energy of the beam particles was {\sl not} used to obtain this signal distribution. The straight line that was used to fit the experimental data points in the scatter plot intersected the $C = S$ line at approximately the correct energy. As is shown in Section VII, this was also true for pions of other energies, for different types of hadrons and also for ``jets", always using the same procedure and the same rotation angle. 

There is no fundamental difference between the way in which the energy resolution is typically measured for other calorimeters and the rotation method described above for a dual-readout calorimeter. The conversion factor between deposited energy (in GeV) and resulting signal (\eg in ADC counts) is established with a beam of mono-energetic electrons. Next, the hadronic energy resolution is determined from the signal distributions measured for beams of mono-energetic hadrons. And unlike in some other calorimeters, no additional information on the beam particles, such as the energy or the hadron type, is used in the rotation method.
 
One may (correctly) argue that the width of signal distributions such as the one shown in Figure \ref{DRrotation}c is not equivalent to the precision with 
which the energy of one arbitrary particle (of unknown energy) absorbed in this calorimeter may be determined. 

%
\begin{figure}[htbp]
\epsfysize=10cm
\centerline{\epsffile{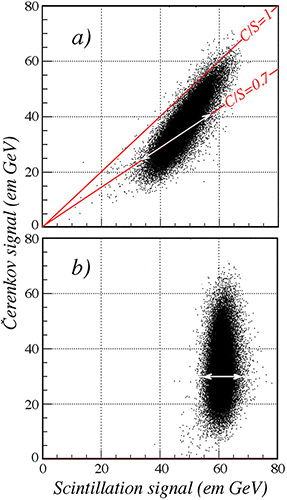}}
\caption{\small{Scatter plots of the \v{C}erenkov \vs the scintillation signals from showers induced by mono-energetic hadrons ($a$). 
The arrow indicates the precision with which the em shower fraction, and thus the energy, of an individual particle can be determined on the basis of the measured ratio of the \v{C}erenkov and scintillation signals, 0.7 in this example. The rotation procedure for an ensemble of 
mono-energetic pions leads to the scatter plot shown in diagram $b$.
The precision of the measurement of the width of that distribution is indicated by a white arrow as well \cite{Lee17}.}}
\label{drmethod}
\end{figure}
To that end, one may use a procedure (described in Section VI), in which
the em shower fraction ($f_{\rm em}$) of the hadronic shower is derived from the ratio of the \v{C}erenkov and scintillation signals. Using the known $e/h$ values of the two calorimeter structures, the measured signals can then be converted to the em energy scale ($f_{\rm em} = 1$). 
The precision obtained with this method is worse than the energy resolution given above. This is mainly due to the fact that contributions of fluctuations in the (very low) \v{C}erenkov light yield, which are responsible for vertical scattering in the $S-C$ diagram, have been eliminated as a result of the rotation. 
Figure \ref{drmethod} graphically illustrates the difference between the values obtained with the two methods discussed here. The precision of the energy measurement is represented by the arrows in the two diagrams \cite{Lee17}.

\section{The DREAM project}

Inspired by the results obtained with the ACCESS calorimeter, discussed in Section IV, the authors embarked on a follow-up project
intended to contain hadron showers in a much more complete way. The instrument they built became known as the DREAM calorimeter.
As before, the two active media were scintillating fibers which measured the visible energy, while clear, undoped fibers measured the generated \v{C}erenkov light. Copper was chosen as the absorber material. 
\begin{figure}[htb]
\epsfysize=10cm
\centerline{\epsffile{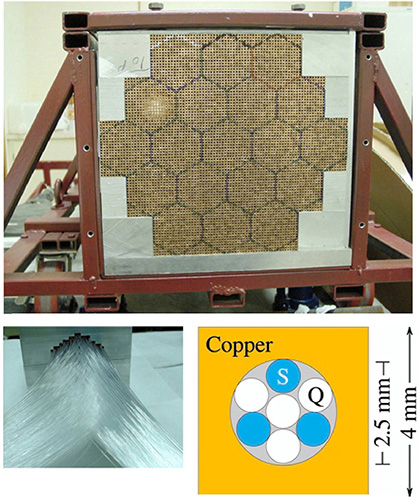}}
\caption{\small
The DREAM detector \cite{Akc05b}. The basic building block is an extruded hollow copper rod. Seven optical
fibers (four \v{C}erenkov and three scintillating fibers) are inserted in the central hole. The
two types of fibers are split into separate bunches as they exit the downstream detector end. The hexagonal readout structure is
indicated. The \v{C}erenkov fibers of the central tower and the six towers of the Inner Ring were filled with quartz fibers, in the twelve towers of the Outer Ring clear PMMA fibers were used for this purpose.}
\label{Detector}
\end{figure}
The basic element of this detector (see Figure \ref{Detector}) was an extruded copper rod, 2 meters long
and 4$\times$4 mm$^2$ in cross section. This rod was hollow, the central cylinder had a diameter of 2.5 mm. 
In this hole were inserted seven optical fibers. Three of these were plastic scintillating fibers, the other four fibers were undoped. 
All fibers had an outer diameter of 0.8 mm and a length of 2.50 m. The fiber pattern was the same for all rods, and is shown in Figure \ref{Detector}.  

The DREAM detector consisted of 5,580 such rods, 5,130 of these were equipped with fibers. 
The empty rods were used as fillers, on the periphery of the detector. The instrumented volume thus had a
length of 2.0 m, an effective radius of $\sqrt{5130\times 0.16/\pi} = 16.2$ cm, and a mass of 1,030 kg. The effective radiation
length ($X_0$) of the calorimeter was 20.1 mm, the Moli\`ere radius ($\rho_M$) was 20.4 mm and the nuclear interaction
length ($\lambda_{\rm int}$) 200 mm.
The composition of the instrumented part of the calorimeter was as follows: 69.3\% of the detector volume consisted of copper
absorber, while the scintillating and \v{C}erenkov fibers occupied 9.4\% and 12.6\%, respectively. Air accounted for the remaining
8.7\%. Given the specific energy loss of a minimum-ionizing particle (mip) in copper (12.6 MeV/cm) and polystyrene (2.00 MeV/cm),
the sampling fraction of the copper/scintillating-fiber structure for mips was thus 2.1\%.     

The fibers were grouped to form 19 towers. Each tower consisted of 270 rods and had an approximately hexagonal
shape (80~mm apex to apex). The layout is schematically shown in Figure \ref{Detector}:
a central tower, surrounded by two hexagonal rings, the Inner Ring (six towers) and the Outer Ring (twelve towers). The towers were
longitudinally unsegmented.

The depth of the copper structure was 200~cm, or 10.0 $\lambda_{\rm int}$. The fibers leaving the rear end of this
structure were separated into bunches: One bunch of scintillating fibers and one bunch of \v{C}erenkov fibers for each tower, 
38 bunches in total. In this way, the readout structure was established (see Figure \ref{Detector}). Each bunch was coupled
through a 2 mm air gap to a photomultiplier tube (PMT). 
\begin{figure}[htb]
\epsfysize=10cm
\centerline{\epsffile{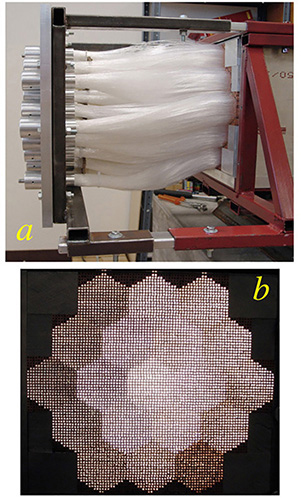}}
\caption{\small
The DREAM calorimeter \cite{Akc05b}. Shown are the fiber bunches exiting from the rear face of the detector ($a$) and a picture 
taken from the front face while the rear end was illuminated ($b$).}
\label{detector2}
\end{figure}

Figure \ref{detector2} shows photographs of the assembled detector. In Figure \ref{detector2}a, the fiber bunches exiting the
downstream end of the calorimeter and the 38 ferrules that hold and position the fibers for the PMTs that detect their signals are shown. In total, this detector contained about
90 km of optical fibers. Figure \ref{detector2}b shows the front face of the calorimeter, when the fibers
were illuminated with a bright lamp located behind the detector. The hexagonal readout structure is clearly visible.
\vskip 2mm
\begin{figure}[htbp]
\epsfysize=14cm
\centerline{\epsffile{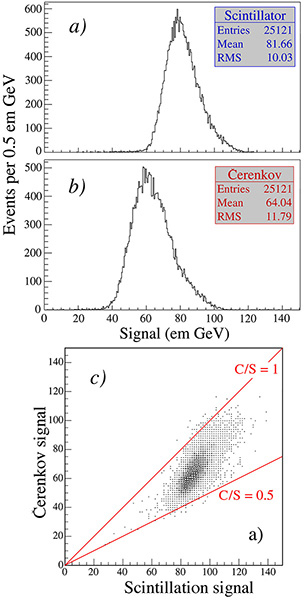}}
\caption{\small
Signal distributions for 100 GeV $\pi^-$ recorded by the scintillating ($a$) and \v{C}erenkov ($b$) fibers of the DREAM
calorimeter, and a scatter plot showing the correlation between both types of signals ($c$). 
The signals are expressed in the same units as those for em showers, which were used to calibrate the calorimeter (em GeV). Data from \cite{Akc05b}.} 
\label{QSbefore}
\end{figure}
Figure \ref{QSbefore} shows the signal distributions for 100 GeV $\pi^-$ detected with this calorimeter. The energy scale was determined with 
electrons, and the average hadronic response was thus 0.8166 for the scintillating fiber structure and 0.6404 for the \v{C}erenkov one. 
The response functions exbibit the asymmetric shape that is characteristic for hadrons in a non-compensating calorimeter (see Figure \ref{femprops}b). 
The correlation between both types of signals is shown in Figure \ref{QSbefore}c. This scatter plot may be compared with the one obtained for the ACCESS calorimeter (Figure \ref{access2}a). The events are now concentrated in a smaller area of the scatter plot, as a result of the better shower containment. However, the fact that the events, as before, are not concentrated along the diagonal, illustrates the complementary information provided by both signals \cite{Akc05b}. 

Using Equation \ref{eq2}, the ratio of the two signals, $C/S$, is related to the em shower fraction, $f_{\rm em}$,  as follows:
\begin{equation}
{C\over S} ~=~ {f_{\rm em} + 0.21~(1 - f_{\rm em})\over {f_{\rm em} + 0.77~ (1 - f_{\rm em})}}
\label{eq6}
\end{equation}
where 0.21 and 0.77 represent the $h/e$ ratios of the \v{C}erenkov and scintillation calorimeter structures, respectively.  
The em shower fraction can thus be determined event-by-event by measuring the $C/S$ signal ratio, and plugging it into Equation \ref{eq3a}.

\begin{figure}[b!]
\epsfysize=14cm
\centerline{\epsffile{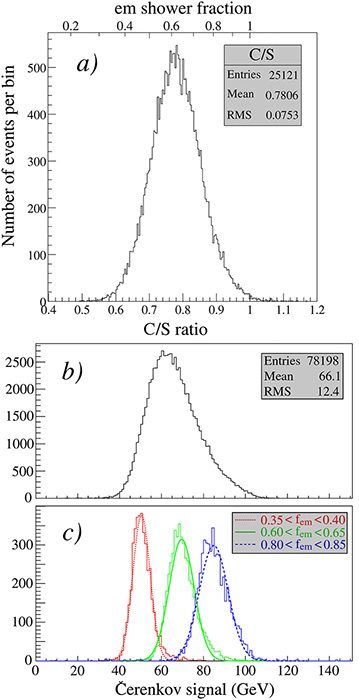}}
\caption{\small
The relationship between the ratio of the \v{C}erenkov and scintillation signals from the DREAM calorimeter and the electromagnetic shower fraction, derived for the 100 GeV $\pi^-$ events on the basis of Equation \ref{eq3} ($a$).
The total \v{C}erenkov signal distribution for these events ($b$) and distributions for subsamples of events selected on the basis of
the measured $f_{\rm em}$ value ($c$). Data from \cite{Akc05b}.}
\label{Pi100C}
\end{figure}
The merits of the dual-readout method are clearly illustrated by Figure \ref{Pi100C} \cite{Akc05b}.
The distribution of the event-by-event signal ratio is shown in Figure \ref{Pi100C}a. The value of $f_{\rm em}$ (top scale) varies from 0.3 to 1, with a maximum around 0.6. The $f_{\rm em}$ value, which can thus be derived from the \v{C}erenkov/scintillation signal ratio for each individual event, can be used to dissect the overall signal distributions. This is illustrated in Figures \ref{Pi100C}b/c, which show  the overall \v{C}erenkov signal
distribution for the 100 GeV $\pi^-$ events, as well as distributions for three subsamples selected on the basis of their $f_{\rm em}$ value. Each $f_{\rm em}$ bin probes a certain region of the overall signal distribution, and the
average value of the subsample distribution increases with $f_{\rm em}$. The overall signal distribution is thus a superposition of many such (Gaussian) subsample signal distributions, and the shape of the overall signal distribution reflects the (asymmetric) distribution of the $f_{\rm em}$ values (see Figure \ref{femprops}b).
\begin{figure}[htb]
\epsfysize=12cm
\centerline{\epsffile{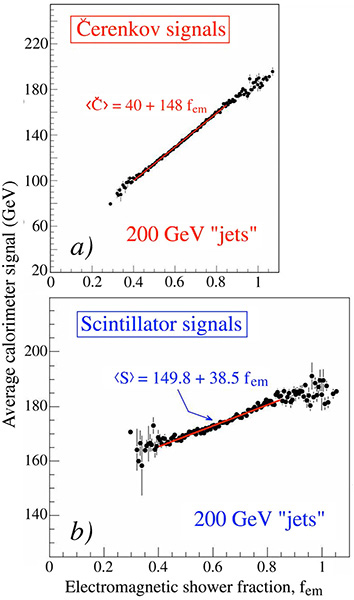}}
\caption{\small
The average \v{C}erenkov ($a$) and scintillation ($b$) signals for 200 GeV ``jets'' in the DREAM calorimeter, as a function of the em shower fraction, $f_{\rm em}$ \cite{Akc05b}.}
\label{Jet200C}
\end{figure}
\begin{figure*}[htbp]
\epsfysize=12.8cm
\centerline{\epsffile{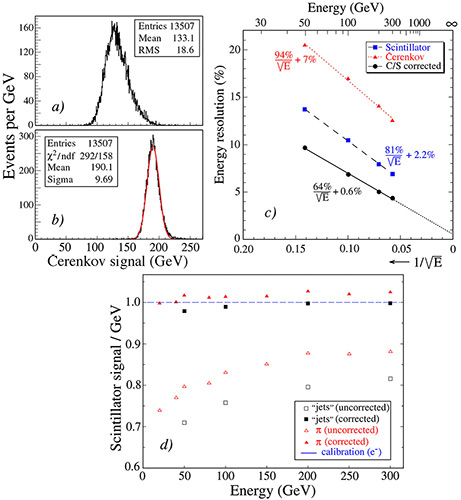}}
\caption{\small Effects of the dual-readout method applied on the basis of the observed \v{C}erenkov/scintillation signal ratio.
\v{C}erenkov signal distributions for high-multiplicity 200 GeV ``jets'' in the DREAM calorimeter before ($a$) and after ($b$) the dual-readout method was applied. The energy resolution for multiparticle ``jets'', measured separately with the scintillation and \v{C}erenkov signals, and after applying the dual-readout method, both for single pions and for multiparticle ``jets''. The electron response, which was the basis for calibrating the calorimeter signals, is shown as well ($d$). Data from \cite{Akc05b}.} 
\label{QSafter}
\end{figure*}

Instead of three $f_{\rm em}$ bins, one could also use a much larger number, and plot the average calorimeter signal as a function of $f_{\rm em}$. The results are shown in Figure \ref{Jet200C} for 200 GeV ``jets'',  separately for the \v{C}erenkov (Figure \ref{Jet200C}a) and scintillation (Figure \ref{Jet200C}b) signals.  
The figure shows linear relationships between these signals and the em shower fraction, thus confirming Equations
\ref{eq3}. These relationships make it possible to determine the $e/h$ values of the calorimeter for the two types of signals.
According to Equation \ref{eq2}, the response should vary between $R = h/e$ for $f_{\rm em} = 0$ and $R = 1$ for $f_{\rm em} = 1$.
The value $R =1$ is obtained based on the assumption that the detected energy was 188 GeV instead of 200, which is reasonable, since some fraction of the particles produced in the upstream pion interactions have not, or only partially been detected by the calorimeter. 
Under that assumption, the fits from Figure \ref{Jet200C} lead to $h/e = 40/188$ for the \v{C}erenkov calorimeter and 
$h/e = 149.8/188$ for the scintillation calorimeter. If one would assume that the entire 200 GeV is deposited in the calorimeter, one would find $e/h$ values of 200/40 = 5.0 and 200/149.8 = 1.34 for the \v{C}erenkov and scintillation calorimeter structures, respectively. These values change to 188/40 = 4.7 and 188/149.8 = 1.26, respectively, under the stated leakage assumption. The inverted values of these ratios, 0.21 and 0.77, are the ones used in Equation \ref{eq6}.

These results may serve to provide a feeling for the experimental uncertainties in the em shower fraction (Equation \ref{eq3}), as well as the energy of the showering hadrons. The latter can be found by solving the two Equations \ref{eq2} for the parameter $E$ (instead of $f_{\rm em}$):
\begin{equation}
E~=~ {{S - \chi C}\over {1 - \chi}},~~~{\rm with}~~\chi = {{1 - (h/e)_S}\over {1 - (h/e)_C}}~\sim 0.3
\label{eq7}
\end{equation}
This expression essentially determines the shower energy by calculating what the calorimeter response would have been for $f_{\rm em} = 1$, based on the actually measured $f_{\rm em}$ value.

A comparison of the scintillation and \v{C}erenkov signals thus made it possible to correct the experimental data in a straightforward way for the effects of non-compensation. In this process, the energy resolution improved, the signal distribution became much more Gaussian and, most importantly, the hadronic energy was correctly reproduced, both for single hadrons and for jets.
The results for 200 GeV ``jets'' are shown in Figure \ref{QSafter}. Using only the
{\em ratio} of the two signals produced by this calorimeter, the resolution for these ``jets''  improved from 14\% to 5\%, in
the \v{C}erenkov channel (Figure \ref{QSafter}a,b). It was shown that this 5\% resolution was in fact dominated by fluctuations in side leakage in this (small, only 1,030 kg instrumented mass) detector. Interestingly, the energy resolution turned out to scale almost perfectly with $E^{-1/2}$ after this $C/S$ information was incorporated (Figure \ref{QSafter}c), while the energy resolution measured for each of the two signals separately showed  large deviations from such scaling. 

Also the jet energy was well reconstructed as a result of this procedure. Whereas the raw data gave a
mean value of 133.1 GeV for these 200 GeV ``jets'', the described procedure led to hadronic energies that were within a few
percent the correct ones, {\em in an instrument calibrated with electrons}. In the process, hadronic signal linearity (a notorious
problem for non-compensating calorimeters) was more or less restored as well (Figure \ref{QSafter}d). 

Monte Carlo simulations indicated that fluctuations in side leakage contributed substantially to the measured hadronic energy resolutions, and would 
most likely be strongly reduced in a sufficiently large detector \cite{Akc14c}.
For example, the energy resolution for 100 GeV pions improved from 7.3\% to 4.6\% when the effective calorimeter radius was doubled.
\begin{figure}[htb]
\epsfxsize=8.7cm
\centerline{\epsffile{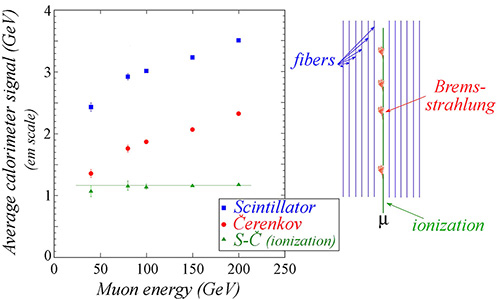}}
\caption{\small Average values of the scintillation and \v{C}erenkov signals from muons traversing the
DREAM calorimeter, as a function of the muon energy. Also shown is the difference between
these signals. All values are expressed in units of GeV, as determined by the electron
calibration of the calorimeter \cite{Akc04}.} 
\label{DREAMmu}
\end{figure}
%

Simultaneous detection of the scintillation and \v{C}erenkov light produced in the shower development turned out to have other, unforeseen beneficial aspects as well. One such effect concerns the detection of muons. Figure \ref{DREAMmu} shows the average signals from muons traversing the DREAM calorimeter along the fiber direction \cite{Akc04}. The gradual increase of the response with the muon energy is a result of the increased contribution of radiative energy loss (bremsstrahlung) to the signals. The \v{C}erenkov fibers are {\em only} sensitive to this energy loss component, since the primary \v{C}erenkov radiation emitted by the muons falls outside the numerical aperture of the fibers. The constant (energy-independent) difference between the total signals observed in the scintillating and \v{C}erenkov fibers thus represents the non-radiative component of the muon's energy loss. Since the signals from both types of fibers were calibrated with em showers, their responses to the radiative component were equal.  This is a unique example of a detector that separates the energy loss by muons into radiative and non-radiative components.
\vskip 2mm

Following the successes of the DREAM project, a new collaboration was formed to explore the possibilities opened up by this new 
calorimeter technique. This became known as the RD52 Collaboration, and its activities were part of the officially supported CERN detector R\&D program. All experimental activities were concentrated in the H8 beam of CERN's Super Proton Synchrotron. In the following section, highlights of the achievements of this project are presented.
\vskip 5mm

\section{The RD52 project}

\subsection{Crystals for dual-readout calorimetry.}
Once the effects of the dominant source of fluctuations are eliminated, the resolution is determined and limited by other types of fluctuations. In the case of the DREAM detector, these fluctuations included, apart from fluctuations in side leakage which can be eliminated by making the detector sufficiently large, {\em sampling fluctuations} and fluctuations in the {\em \v{C}erenkov light yield}. The latter effect alone contributed $35\%/\sqrt{E}$ to the measured resolution, since the quartz fibers generated only eight \v{C}erenkov photoelectrons per GeV deposited energy. 
Both effects could in principle be greatly reduced by using crystals for dual-readout purposes. Certain dense high-$Z$ crystals (PbWO$_4$, BGO) produce significant amounts of \v{C}erenkov light.
The challenge is of course to separate this light effectively from the (overwhelmingly) dominant scintillation light.
Precisely for that reason, the idea to use such crystals as dual-readout calorimeters met initially with considerable doubt. Yet, the RD52 Collaboration demonstrated that it could be done.

For the proof-of-principle measurements, lead tungstate (PbWO$_4$) crystals were used. This material has the advantage of producing relatively very little scintillation light, while the large refractive index promised a substantial \v{C}erenkov light yield.
\v{C}erenkov light is emitted by charged particles traveling faster than $c/n$, \ie the speed of light in the medium with refractive index $n$ in which this process takes place. The light is emitted at a characteristic angle, $\theta_C$, defined by $\cos \theta_C = 1/\beta n$.
When sufficiently relativistic particles (\ie $\beta \sim 1$) traverse PbWO$_4$ crystals ($n = 2.2$), $\theta_C \sim 63^\circ$~\footnote{The reality may be somewhat more complicated, because of the 
anisotropic optical properties of lead tungstate crystals \protect\cite{Bac97,Chi00}, which might affect some aspects of \v{C}erenkov light emission \protect\cite{Del98}.}. 

\begin{figure}[htbp]
\epsfxsize=8.7cm
\centerline{\epsffile{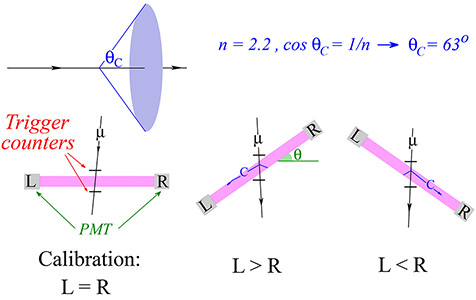}}
\caption{\small
Principle of the asymmetry measurement that was used to establish the contribution of \v{C}erenkov light to the 
signals from the PbWO$_4$ crystals. Depending on the crystal orientation, this directionally emitted light contributed differently to the signals from the left and right photomultiplier tubes \cite{Akc07b}. }
\label{chproof1}
\end{figure}

In order to detect the contribution of \v{C}erenkov light to the signals from a PbWO$_4$ crystal, 
both ends of the crystal were equipped with a photomultiplier tube. By varying the detector {\sl orientation} with respect to the direction of the incoming particles, 
a contribution of \v{C}erenkov light would then manifest itself as an angle-dependent asymmetry. 
This is illustrated in Figure \ref{chproof1}, which shows the setup of the initial measurements that were performed with a cosmic-ray telescope to test this principle \cite{Akc07b}. The PMT gains were equalized
for the leftmost geometry, in which the crystal was oriented horizontally. 
By tilting the crystal through an angle ($\theta$) such that the axis of the crystal is oriented at the \v{C}erenkov angle 
$\theta_C$ with respect to the particle direction, \v{C}erenkov light produced by the cosmic rays traversing the trigger counters would be preferably detected in either the $L$ (central geometry) or $R$ (rightmost geometry)
PMT. By measuring the response asymmetry $(R-L)/(R+L)$ as a function of the tilt angle $\theta$, the contribution of \v{C}erenkov light to the detector signals could be determined.
\begin{figure}[htbp]
\epsfxsize=8.7cm
\centerline{\epsffile{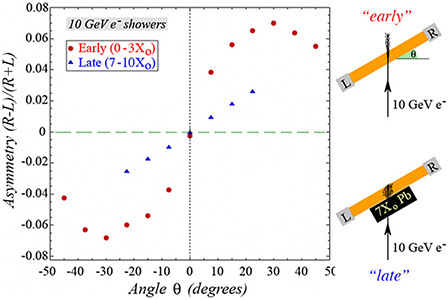}}
\caption{\small
Left-right response asymmetry measured for 10 GeV electrons showering in a $2.5 X_0$ thick 
PbWO$_4$ crystal, as a function of the orientation of the crystal (the angle $\theta$). Results are shown for the early and the late components of the showers. The latter measurements were done by placing 4 cm of lead upstream of the crystal \cite{Akc07b}. }
\label{chproof2}
\end{figure}

The initial cosmic-ray measurements indicated that the contribution of \v{C}erenkov light to the signals was at the level of 15 -- 20\% \cite{Wig07}. Because of the extremely low event rates and the tiny signals (typically 20 -- 30 MeV), systematic follow-up studies were carried out with particle beams at CERN's SPS. 
Figure \ref{chproof2} shows some of the results of this work, and in particular the characteristic ``S'' shape which indicates that the \v{C}erenkov component of the light produced in the developing showers was most efficiently detected when the crystal axis was oriented at the \v{C}erenkov angle with the shower axis. The figure also shows that placing a lead brick upstream of the crystal had the effect of making the angular distribution of the light produced in the crystal more isotropic, thus reducing the left-right asymmetry \cite{Akc07b}.

It turned out that the scintillation light yield, and thus the fraction of \v{C}erenkov light in the overall signal, depends very sensitively on the temperature of the PbWO$_4$ crystals, approximately -3\% per degree Celsius \cite{Akc08b}. For this reason, the large em calorimeters that are based on these crystals (CMS, ALICE, PANDA) all operate at very low temperatures, and maintaining the temperature constant at the level of $\pm 0.1 ^\circ$C is an essential requirement for obtaining excellent energy resolution. There is also another temperature dependent phenomenon that affects the efficiency at which the \v{C}erenkov and scintillation components of the light produced by these crystals can be separated, namely the decay time of the scintillation signals. RD52 found this decay time to decrease from $\sim 9$ ns at $13^\circ$C to $\sim 6$ ns at $45^\circ$C \cite{Akc08b}. 
\begin{figure}[b!]
\epsfysize=6.5cm
\centerline{\epsffile{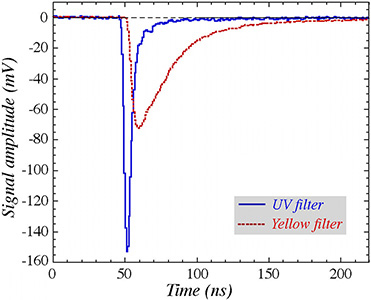}}
\caption{\small
Average time structure of the signals from a PbWO$_4$ crystal doped with 1\% Mo, generated by 50 GeV electrons. The angle $\theta$ was 30$^\circ$ in these measurements.
Shown are the results obtained with UV and yellow filters, respectively \cite{Akc09b}.}
\label{Modoping}
\end{figure}

The difference in the time structure of the two signals is another important characteristic that can be used to distinguish between the scintillation and \v{C}erenkov components of the light produced by high-energy particles in crystals. And of course, the larger the difference in the time structure, the better the separation works. The RD52 collaboration managed to improve the applicability of PbWO$_4$ crystals for dual-readout calorimetry by doping them with small amounts, $\cal{O}$(1\%), of molybdenum \cite{Akc09b}. This had two beneficial effects: it increased the decay time of the scintillation light and it shifted the spectrum of the emitted scintillation light to larger wavelengths.

The effects of that are illustrated in Figure \ref{Modoping}, which shows the calorimeter signals generated by 50 GeV electrons traversing a crystal of this type. This crystal was oriented such as to maximize the relative fraction of \v{C}erenkov light in the detected signals. By selecting the UV light by means of an optical filter, almost the entire detected signal was due to (prompt) \v{C}erenkov light, while a yellow transmission filter predominantly selected scintillation light, which had a decay time of $\sim 26$ ns as a result of the Mo-doping. 

Whereas the differences in angular dependence were very suitable for demonstrating the fact that some of the light generated in these crystals is actually the result of the \v{C}erenkov mechanism, the combination of time structure and spectral characteristics provides powerful tools to separate the two types of light in real time. One does not even have to equip the calorimeter with two different light detectors for that. This was demonstrated with a calorimeter consisting of bismuth germanate (Bi$_4$Ge$_3$O$_{12}$, or BGO) crystals \cite{Akc09c}. 
Even though \v{C}erenkov radiation represents only a very tiny fraction of the light produced by these crystals, it is relatively easy to separate and extract it from the signals. The much longer scintillation decay time (300 ns) and the spectral difference are responsible for that \footnote{The BGO scintillation spectrum peaks at 480 nm, while \v{C}erenkov light exhibits a $\lambda^{-2}$ spectrum.}. 
\begin{figure}[htb]
\epsfysize=7cm
\centerline{\epsffile{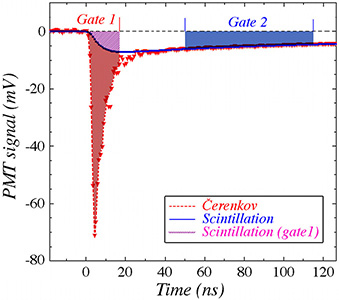}}
\caption{\small
The time structure of a typical shower signal from 50 GeV electrons measured in the BGO em calorimeter equipped with a UV filter. These signals were measured with a sampling oscilloscope, which took a sample every 0.8 ns \cite{Akc09c}. The UV signals were used to measure the relative contributions of scintillation light (gate 2) and \v{C}erenkov light (gate 1).}
\label{BGOtime}
\end{figure}

Figure \ref{BGOtime} shows the time structures of signals from a BGO calorimeter recorded with a UV filter.
The ``prompt" component observed in the ultraviolet signal is due to \v{C}erenkov light. A small fraction of the scintillation light also passes through the UV filter.
This offers the possibility to obtain all needed information from only one signal. 
An external trigger opens two gates: one narrow (10 ns) gate covers the prompt component, the second gate (delayed by 30 ns and 50 ns wide) only contains scintillation light. The latter signal can also be used to determine the contribution of scintillation to the light collected in the narrow gate. In this way, the \v{C}erenkov/scintillation ratio can be measured event-by-event on the basis of one signal only \cite{Akc09c}. 

\begin{figure}[htb]
\epsfysize=11cm
\centerline{\epsffile{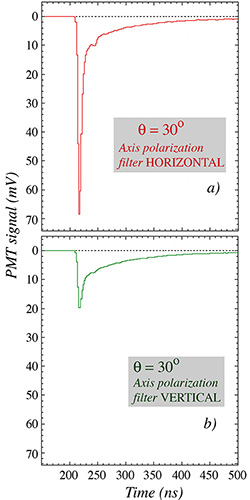}}
\caption{\small
Average time structure of the signals generated by 180 GeV $\pi^+$ traversing a BSO crystal in its center at $\theta = 30^\circ$ and passing through a U330 optical transmission filter, followed by a polarization filter. The transmission axis of the latter filter was either oriented horizontally ($a$) or vertically ($b$). The time scale describes the time passed since the start of the time base of the oscilloscope \cite{Akc11a}.}
\label{Polarization}
\end{figure}

The same possibility was offered by BSO crystals. These have a similar chemical composition as BGO, with the germanium atoms replaced by silicon ones. 
Both the (scintillation) light yield and the decay time of this crystal are about a factor of three smaller than for BGO. Tests with BSO crystals showed that this made the separation of \v{C}erenkov and scintillation light somewhat more efficient, while maintaining the possibility to obtain all necessary information from one calorimeter signal. This, combined with the fact that the expensive germanium component is not needed, makes BSO a potentially interesting candidate for a crystal-based dual-readout calorimeter \cite{Akc11b}.
\vskip 2mm
Apart from the time structure and the spectral differences, there is one other characteristic feature of \v{C}erenkov light that can be used to distinguish it from scintillation light, namely the fact that it is {\sl polarized} \cite{Akc11a}. The polarization vector is oriented perpendicular to the surface of the cone of the emitted \v{C}erenkov light. RD52 used a BSO crystal to demonstrate this possibility (Figure \ref{Polarization}). 
This crystal was placed in a particle beam and oriented such as to maximize the fraction of \v{C}erenkov light that reached the PMT (as in Figure \ref{chproof2}). 
A UV filter absorbed most of the scintillation light, and the time structure of the transmitted signals showed a very significant prompt \v{C}erenkov signal, as well as a 100 ns tail due to the transmitted component of the scintillation light. In addition, a polarization filter was placed directly in front of the PMT. Rotating this filter over 90$^\circ$ had a major effect on the prompt \v{C}erenkov component, while the scintillation component was not affected at all \cite{Akc11a}.

\subsection{Tests of crystal-based dual-readout calorimeters.}

The RD52 collaboration also performed tests of calorimeter systems in which the em section consisted of high-$Z$ crystals, while the original DREAM fiber calorimeter served as the hadronic section \cite{Akc09c}. Two matrices of crystals were assembled for this purpose. The first one consisted of 19 PbWO$_4$ crystals borrowed from the CMS Collaboration (total mass $\sim 20$ kg). 
\begin{figure}[htb]
\epsfysize=7.5cm
\centerline{\epsffile{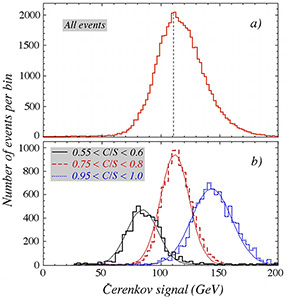}}
\caption{\small
The \v{C}erenkov signal distribution for 200 GeV ``jet" events detected in the BGO+fiber calorimeter system ($a$), together with the distributions for subsets of events selected on the basis of the ratio of the total \v{C}erenkov and scintillation signals in this detector combination ($b$). Data from \cite{Akc09c}.}
\label{BGOmatrix}
\end{figure}
The second matrix consisted of 100 BGO crystals that were previously used in the em calorimeter of the L3 experiment (total mass $\sim 150$ kg) \cite{Sum88}. The \v{C}erenkov and scintillation components of the light produced in these crystals were separated as described in the previous subsection, exploiting the differences in time structure and spectral composition.
 Figure \ref{BGOmatrix} shows results from the measurements with the BGO matrix, obtained for high-multiplicity multiparticle events (``jets'') generated by 200 GeV $\pi^+$ in an upstream target. The overall \v{C}erenkov signal distribution is shown, together with subsets of events selected on the basis of the measured \v{C}erenkov/scintillation signal ratio, \ie on the basis of the em shower content of the events. A comparison with Figure \ref{Pi100C} indicates that the dual-readout method also worked for this detector combination \cite{Akc09c}.
\vskip 2mm
\begin{figure*}[htbp]
\epsfysize=7cm
\centerline{\epsffile{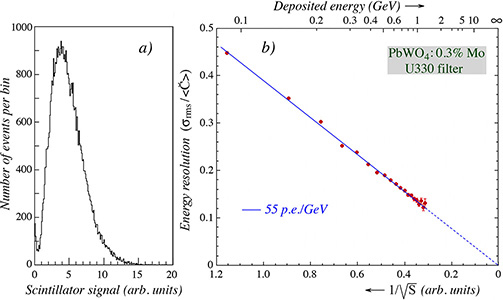}}
\caption{\small
The scintillation signal distribution for 50 GeV electrons traversing a PbWO$_4$ crystal at $\theta =30^\circ$ ($a$)
and the fractional width of the \v{C}erenkov signal distribution as a function of the amount of energy deposited in the crystal, as derived from the scintillation signal ($b$). The crystal was doped with 0.3\% Mo \cite{Akc10}. See the text for more details. }
\label{Cyield}
\end{figure*}

Yet, after elaborate studies of many crystals, and dedicated efforts to tailor the crystal properties to the specific requirements for dual-readout calorimetry, the RD52 Collaboration decided that this was not the most promising avenue for improving the performance obtained with the original DREAM calorimeter. We recall that the main motivation for examining the option of using crystals for dual-readout calorimetry was the possibility to eliminate the effects of sampling fluctuations, and the potentially higher \v{C}erenkov light yield, thus reducing the main sources 
of fluctuations that limited the performance of the DREAM fiber calorimeter. However, it turned out that the use of crystals introduced new, worse sources of fluctuations. 

The main problem is the fact that the (short wavelength) light that constitutes the \v{C}erenkov signals is strongly attenuated, because of the absorption characteristics of the crystals. The attenuation length was in some cases so short that it led to $\cal{O}$(10\%) response non-linearities for electron showers \cite{Akc12b}. Since the depth at which the light is produced increases only logarithmically with the electron energy, this indicates that the attenuation length is of the order of a few radiation lengths.
Such a short attenuation length affects several aspects of the calorimeter performance in major ways, since it causes the signal to depend sensitively 
on the location where the light is produced. For comparison, we mention that the attenuation lengths of the fibers used in the 
dual-readout fiber calorimeters were orders of magnitude longer. In some cases, $\lambda_{\rm att}$ was measured to be more than 20 meters.
Another problem is the fact that a large fraction of the potentially available \v{C}erenkov photons needs to be sacrificed in order to extract a sufficiently pure \v{C}erenkov signal from the light produced by the crystals.

Figure \ref{Cyield} illustrates how this \v{C}erenkov light yield can be measured in practice \cite{Akc10}.
It concerns measurements on a PbWO$_4$ crystal doped with 0.3\% of molybdenum. This crystal was placed at an angle $\theta = 30^\circ$ with the beam line (as in Figure \ref{chproof2}). One PMT ($R$) was equipped with a UV filter, in order to select the \v{C}erenkov light, for which the detection efficiency is largest at this angle. At the other side of the crystal only scintillation light was detected. EGS4 \cite{Nel78} calculations indicated that the  beam particles (50 GeV electrons) deposited on average 0.578 GeV in this crystal, which was slightly thicker than $2X_0$ in this geometry. This made it possible to calibrate the scintillation signals, the distribution of which is shown in Figure \ref{Cyield}a. This distribution was subdivided into 20 bins. For each bin, the signal distribution on the opposite side of the crystal, \ie the \v{C}erenkov side, was measured. The fractional width of this distribution, $\sigma_{\rm rms}/C_{\rm mean}$ , is plotted in Figure \ref{Cyield}b versus the average scintillation signal in this bin, or rather versus the inverse square root of this signal ($S^{-1/2}$). It turned out that this fractional width scaled perfectly with this variable, \ie with $E^{-1/2}$. Since the relationship between the energy $E$ and the scintillation signal $S$ is given by the calibration described above, it was also possible to indicate the energy scale in Figure \ref{Cyield}b. This is done on the top horizontal axis.
The fact that $\sigma_{\rm rms}/C_{\rm mean}$ scales with $E^{-1/2}$ means that the energy resolution is completely determined by stochastic processes that obey Poisson statistics. In this case, fluctuations in the \v{C}erenkov light yield were the only stochastic processes that played a role, and therefore the average light yield could be directly determined from this result: 55 photoelectrons per GeV deposited energy. For an energy deposit of 1 GeV, this led to a fractional width of 13.5\%, and therefore the contribution of \v{C}{erenkov photoelectron statistics amounts to $13.5\%/\sqrt{E}$. This is not much better than what could be achieved in a dedicated fiber sampling calorimeter.

Other considerations that led to the decision to pursue other alternatives for improving the DREAM results were the high cost of the crystals, as well as the fact that the short-wavelength light needed to extract the \v{C}erenkov signals made the crystals very prone to radiation damage effects. Also the requirement to integrate the signals over relatively long time intervals to separate the two types of signals, as illustrated in Figures \ref{Modoping} and \ref{BGOtime}, is a disadvantage in experiments where fast signals are needed, for example to mitigate pileup effects. We recall that the absence of the need for long signal integration times, a key aspect of compensating calorimeters, was one of the reasons to investigate the possibilities of dual-readout calorimetry (Section III).

The alternative chosen by the RD52 Collaboration was the improvement of dual-readout fiber calorimetry, by constructing a very-fine-sampling device, that became known as the SuperDREAM calorimeter. However, before describing this device in detail, one other aspect of high-precision calorimetry is discussed: neutron detection. 

\subsection{Benefits of neutron detection}

Should one succeed to eliminate, or at least greatly reduce the contributions of sampling fluctuations and photoelectron statistics to the hadronic energy resolution, then the last hurdle toward ultimate performance is formed by the fluctuations in invisible energy, \ie fluctuations in the 
energy fraction that is used to break up atomic nuclei. The elimination of fluctuations in $f_{\rm em}$, which can be achieved with dual-readout calorimetry, takes care of the effects of the {\em average} contribution of invisible energy. However, for a given value of $f_{\rm em}$,
the invisible energy fluctuates around this average. The kinetic energy carried by the neutrons produced in the shower development process is correlated to this invisible energy loss
(Figure \ref{brau}). 
\begin{figure}[htbp]
\epsfysize=9cm
\centerline{\epsffile{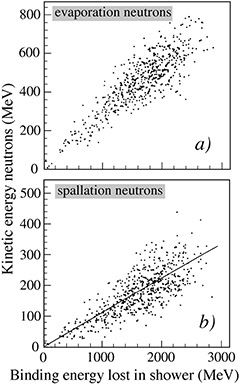}}
\caption{\small
Scatter plots showing the correlation between the kinetic energy carried by soft neutrons ($E < 20$
MeV) and the nuclear binding energy lost when 5 GeV $\pi^-$ mesons are absorbed in depleted uranium
($^{238}$U). The distributions are shown separately for neutrons that originated from evaporation
processes ($a$) and from nuclear spallation ($b$). Results from Monte Carlo simulations with the
HETC/MORSE package by Brau and Gabriel \cite{Brau89}.}
\label{brau}
\end{figure}
Efficient neutron detection can not only reduce the $e/h$ ratio to 1.0, but it also greatly reduces the contribution of fluctuations in invisible energy to the hadronic energy resolution. The practical importance of this is illustrated by the fact that the hadronic energy resolution of compensating calorimeters based on liquid-argon (which is rather insensitive to neutrons) as active medium never reached the good values obtained with plastic-scintillator readout. 
\begin{figure}[htb]
\epsfysize=13cm
\centerline{\epsffile{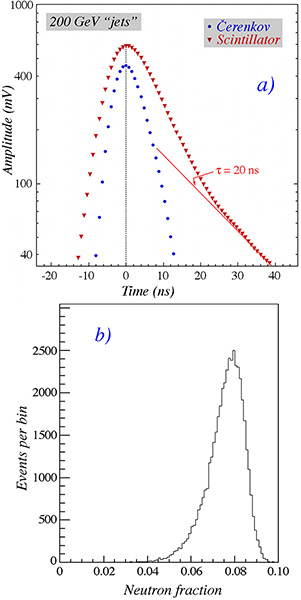}}
\caption{\small The average time structure of the \v{C}erenkov  and  scintillation signals measured for the showers from 200 GeV ``jets"
in the DREAM tower located on the shower axis. The measured (oscilloscope) signals have 
been inverted ($a$). Event-by-event distribution of the fraction of this scintillation signal attributed to neutrons ($b$). Data from \cite{Akc09a}.} 
\label{SQ11}
\end{figure}
It has been demonstrated that efficient neutron detection reduces the ultimate limit on this resolution to $(13.4 \pm 4.7\%)/\sqrt{E}$, in compensating lead/plastic-scintillator calorimeters \cite{Dre90}.

Detailed measurements of the time structure of the calorimeter signals, examples of which are given in Figures \ref{Modoping} and \ref{BGOtime}, make it also possible to measure the contribution of neutrons to the shower signals. Figure \ref{SQ11}a illustrates this with data taken with the original DREAM fiber calorimeter \cite{Akc09a}. The figure shows the average time structure of \v{C}erenkov and scintillation signals measured with a sampling oscilloscope for  showers from 200 GeV multiparticle events developing in this calorimeter.
The scintillation signals exhibit an exponential tail with a time constant of $\sim 20$ ns. This tail has all the characteristics expected of a (nonrelativistic) neutron signal and was absent in the time structure of the \v{C}erenkov signals. It was also not observed in the scintillation signals for em showers \cite{Akc07a}.
The distribution of the contribution of this tail to the hadronic scintillation signals ($f_n$) is plotted in Figure \ref{SQ11}b. 
By measuring the contribution of this tail {\em event by event}, the hadronic energy resolution could be further improved \cite{Akc09a}.
\begin{figure}[htbp]
\epsfysize=12cm
\centerline{\epsffile{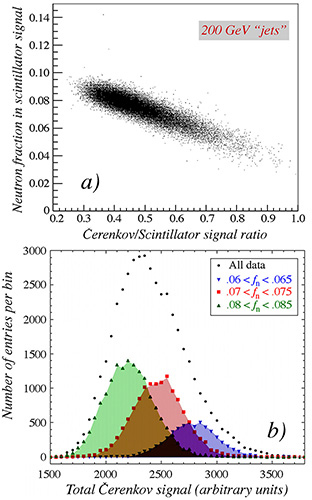}}
\caption{\small Scatter plot for 200 GeV ``jets" in the DREAM calorimeter. For each event, the combination of the total \v{C}erenkov/scintillation signal ratio and the fractional contribution of neutrons to the total scintillation signal is represented by a dot ($a$). Distribution of the total \v{C}erenkov signal for 200 GeV ``jets" and the distributions for three subsets of events selected on the basis of the fractional contribution of neutrons to the scintillation signal ($b$).
Data from \cite{Akc09a}.} 
\label{neutron1}
\end{figure}

It was found that the fraction of the scintillation signal that could be attributed to neutrons ($f_n$) was anti-correlated with the 
\v{C}erenkov/scintillation signal ratio, and thus with $f_{\rm em}$ (Figure \ref{neutron1}a). This is of course no surprise, since a large em shower fraction implies that a relatively small fraction of the shower energy has been used for the processes in which atomic nuclei are broken up. This anti-correlation means that the essential advantages of the dual-readout method, which derived from the possibility to measure $f_{\rm em}$ event by event, could also be achieved with {\em one readout medium}, provided that  the time structure of the (scintillation) signals is measured in such a way that the contribution of neutrons can be determined event by event.

This is illustrated in Figure \ref{neutron1}b, which shows the 
total \v{C}erenkov signal distribution for all 200 GeV ``jet" events, as well as the distributions for subsamples of events with $0.06<f_n<0.065$ (the blue downward pointing triangles), $0.07<f_n<0.075$ (red squares) and $0.08<f_n<0.085$ (green upward pointing triangles). Clearly, the different subsamples each probe a different region of the total signal distribution for all events. This total \v{C}erenkov signal distribution for all events is thus a superposition of many distributions such as the ones for the subsamples shown in this figure. Each of these distributions for the subsamples has a different mean value, and a resolution that is substantially better than that of the overall signal distribution. The signal distributions for the subsamples are also much more Gaussian than the overall signal distribution, whose shape is simply determined by the extent to which different $f_n$ values occurred in practice. And since the $f_n$ distribution is skewed to the low side (Figure \ref{SQ11}b), the overall \v{C}erenkov signal distribution is skewed to the high side.
\begin{figure}[b!]
\epsfysize=12.5cm
\centerline{\epsffile{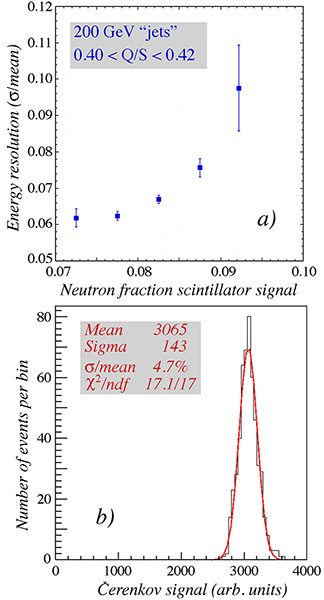}}
\caption{\small The energy resolution measured for the \v{C}erenkov signals from 200 GeV ``jets" with the same em shower fraction, as a function of the fractional neutron contribution to the scintillation signals ($a$). \v{C}erenkov signal distribution for 200 GeV ``jets" with $0.70<C/S<0.75$ and $0.045<f_n<0.065$, together with the results of a Gaussian fit ($b$). Experimental data obtained with the DREAM calorimeter \cite{Akc09a}.} 
\label{neutron2}
\end{figure}

A measurement of the relative contribution of neutrons to the hadronic scintillation signals thus offers similar possibilities for eliminating the effects of non-compensation as an event-by-event measurement of the em shower fraction (Figure \ref{Pi100C}). However, when both $f_{\rm em}$ and $f_n$ are being measured, even better results may be expected.
By selecting a subsample of hadronic events, all with the same $f_{\rm em}$ value, there would still be event-by-event differences in the share of invisible energy.
The nuclear reactions taking place in the non-em shower development process vary from event to event, and so does the nuclear binding energy lost in these processes. For this reason, in calorimeters such as the DREAM fiber calorimeter, measurements of $f_n$ provide information {\em complementary} to that obtained from the $C/S$ signal ratio \cite{Akc09a}.

This is illustrated in Figure \ref{neutron2}a, which shows that the energy resolution of a sample of events with the same em shower fraction is clearly affected by the relative contribution of neutrons to the signals. As $f_n$ increases, so does the fractional width of the \v{C}erenkov signal distribution. A larger $f_n$ value means that the average invisible energy fraction is larger. This in turn 
implies that the event-to-event fluctuations in the invisible energy are larger, which translates into a worse energy resolution, even for signals to which the neutrons themselves do not contribute. Figure \ref{neutron2}b shows the response function obtained with the combined information on the em shower fraction and the contribution of neutrons to the signals. This \v{C}erenkov signal distribution concerns 200 GeV ``jet'' events with a $C/S$ value between 0.70 and 0.75 and a fractional neutron contribution to the scintillation signals between 0.045 and 0.065. The distribution is 
very well described by a Gaussian fit, with an energy resolution of 4.7\%, which may be compared with the 5.1\% resolution obtained when only information on $f_{\rm em}$ was used (Figure \ref{QSafter}b).
The resolution was further reduced, to 4.4\%, when the relative neutron contribution was narrowed down to 0.05 - 0.055. Note that these results were achieved for a calorimeter with an instrumented mass of only about one ton.

\subsection{The RD52 fiber calorimeter}

The design of the new dual-readout fiber calorimeter was driven by the desire to reduce the factors that limited the hadronic energy resolution of the original DREAM fiber calorimeter as much as possible. These factors concerned side leakage,  the \v{C}erenkov light yield and sampling fluctuations. The fluctuations in side leakage could be reduced in a trivial way, \ie by making the calorimeter sufficiently large. It was estimated that the instrumented mass has to be about 5,000 kg to contain hadronic showers at the 99\% level, and thus limit the effects of leakage fluctuations on the hadronic energy resolution to $\sim 1\%$.

Fluctuations in the number of \v{C}erenkov photons would be limited by maximizing the \v{C}erenkov light yield. This could be achieved by
\begin{itemize}
\item Increasing the numerical aperture of the \v{C}erenkov fibers, and/or
\item Aluminizing the upstream end of the \v{C}erenkov fibers, and/or
\item Increasing the quantum efficiency of the PMT photocathodes,
\end{itemize}
while sampling fluctuations can be limited in the following way:
\begin{itemize}
\item Fibers are individually embedded in the absorber structure, instead of in groups of seven,
\item The packing fraction of the fibers is maximized, \ie roughly doubled compared to the DREAM calorimeter.
\end{itemize}
\begin{figure}[htbp]
\epsfxsize=8.7cm
\centerline{\epsffile{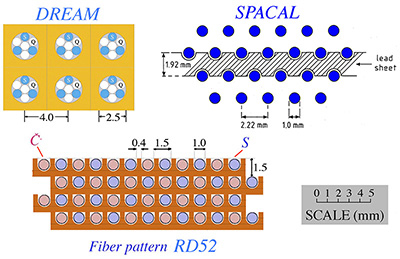}}
\caption{\small The structure of the RD52 fiber calorimeter (copper-based modules), compared to that of two other fiber calorimeters: DREAM \cite{Akc05b} and SPACAL \cite{Aco91c}. From reference \cite{Akc14b}.}
\label{dreamstructure}
\end{figure}
The fiber structure of the RD52 calorimeter is schematically shown in Figure \ref{dreamstructure}. On the same scale, the structures of the DREAM and SPACAL calorimeters are shown as well. Compared with DREAM, the number of fibers per unit volume, and thus the sampling fraction, is approximately twice as large in the RD52 calorimeter. And since each fiber is now separately embedded in the absorber structure, the sampling {\em frequency} has also considerably increased. 
\begin{figure}[htb]
\epsfysize=10cm
\centerline{\epsffile{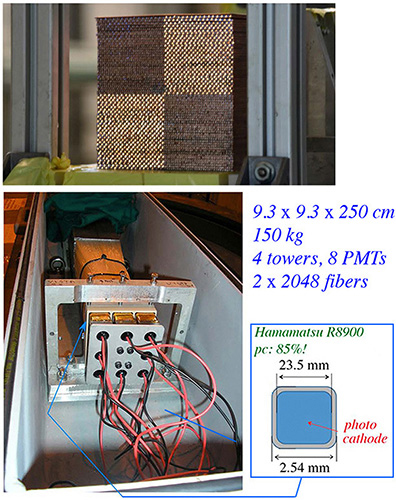}}
\caption{\small Front (left) and rear (right) view of one of the RD52 fiber calorimeter modules. The tower structure is made visible by shining light on two of the eight fiber bunches sticking out at the back end. See text for more details.}
\label{rd52calo}
\end{figure}
Since both factors determine the electromagnetic energy resolution, one should thus expect a substantial improvement (see Equation \ref{eq:reso}). 
\begin{figure*}[htbp]
\epsfysize=8cm
\centerline{\epsffile{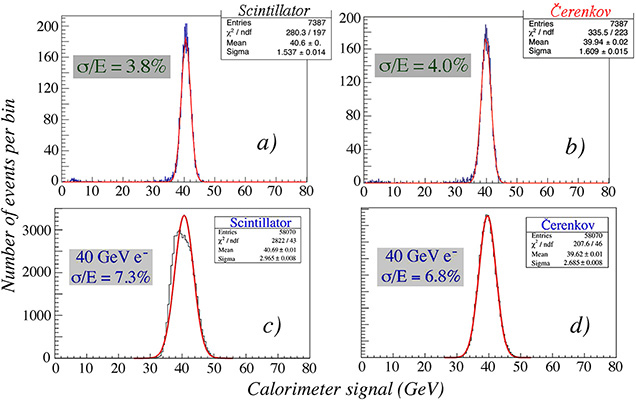}}
\caption{\small Comparison of the em response functions measured with the RD52 copper-fiber calorimeter ($a,b$) \cite{Akc14b} and the 
original DREAM copper-fiber calorimeter ($c,d$) \cite{Akc05a}, for 40 GeV electrons. The angle of incidence of the electron beam with the fiber axis $(\theta, \phi)$ was $(1.5^\circ, 1.0^\circ)$ for RD52 and $(3^\circ, 2^\circ)$ for DREAM.
Results are given separately for the scintillation and \v{C}erenkov signals. }
\label{rd52em1}
\end{figure*}

Figure \ref{rd52calo} shows pictures of the front face and the back end of a calorimeter module. Each module consists of four towers, and each tower produces a scintillation and a \v{C}erenkov signal. The transverse dimension of the module was chosen such that the eight PMTs would fit within its perimeter, and the maximum fiber density was determined by the total photocathode surface of these PMTs (which corresponds to
more than half of the module's lateral cross section).

The \v{C}erenkov light yield was increased by using clear plastic fibers instead of the quartz ones used in the DREAM calorimeter. The numerical aperture of these  plastic fibers is larger (0.50 \vs 0.33) \footnote{The light yield is proportional to the numerical aperture {\em squared}.}. Also, the \v{C}erenkov fiber density was increased, by $\approx 65\%$.
In addition, the new PMTs have a higher quantum efficiency, thanks to a Super Bialkali photocathode. As a result, the number of \v{C}erenkov photoelectrons measured for em showers increased by about a factor of four, from 8 to 33 Cpe/GeV \cite{Akc14b}.

Another important difference between the RD52 and DREAM fiber calorimeters concerns the readout, which in the RD52 one is based on a 
Domino Ring Sampler (DRS) circuit \cite{Rit10} that allows time structure measurements of each signal with a sampling rate of 5 GHz (\ie 0.2 ns time bins). In the previous subsections it was shown  that detailed measurements of the time structure are an invaluable source of information, 
not only for separating the \v{C}erenkov and scintillation signals from crystals, but also to identify and measure the contribution of neutrons to the scintillation signals \cite{Akc09a}. Another important goal
of the time structure measurements is to determine the depth at which the light is produced in this longitudinally unsegmented calorimeter. As is shown in Section VII.E.2,  this can be achieved by making use of the fact that the light signals travel at a slower speed in the fibers ($\sim 17$~cm/ns) than the particles producing this light ($30$~cm/ns).

It turned out to be very difficult to produce copper plates with the required specifications for this very-fine-sampling calorimeter structure.
Therefore, the collaboration initially built nine modules using lead, which is relatively easy to extrude, as the absorber material. At a later stage, also several copper modules were built.

\subsubsection{\sl Electromagnetic performance.}

The RD52 calorimeter modules were extensively tested with beams of electrons, with energies ranging from 6 - 80 GeV.
For reasons discussed in Section VII.E.1, the scintillation resolution turned out to be very sensitive to the angle of incidence of the particles, when these angles were very small ($< 3^\circ$ between the beam line and the direction of the fibers) and the electron energy was high.
Figure \ref{rd52em1} shows the response functions for 40 GeV electrons, separately for the scintillation and the \v{C}erenkov signals \cite{Akc14b}. For comparison, the response functions measured with the original DREAM fiber calorimeter are shown as well. The energy resolution was considerably better, and the response functions were also better described by a Gaussian function, especially in the case of the scintillation signals. This despite the fact that the RD52 measurements were performed at a much smaller angle of incidence: $(\theta, \phi) = (1.5^\circ, 1.0^\circ)$ \vs $(3^\circ, 2^\circ)$ for DREAM \cite{Akc05a}.
\begin{figure}[b!]
\epsfysize=6cm
\centerline{\epsffile{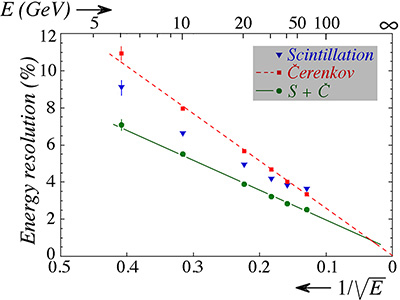}}
\caption{\small The em energy resolution measured with the \v{C}erenkov fibers, the scintillating fibers and the sum of all fibers in the RD52 copper-fiber calorimeter \cite{Akc14b}. }
\label{rd52em2}
\end{figure}
\begin{figure}[b!]
\epsfysize=9.8cm
\centerline{\epsffile{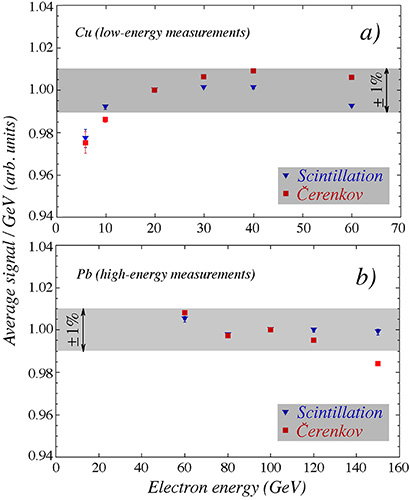}}
\caption{\small The signal linearity for electron showers, measured for the RD52 copper ($a$) and lead ($b$) modules \cite{Akc14b}. }
\label{linearity_em}
\end{figure}

One advantage of the new fiber pattern used in the RD52 calorimeters is the fact that the scintillation and \v{C}erenkov readout represent completely independent sampling structures. Therefore, by combining the signals from the two types of fibers, a significant improvement in the energy resolution could be obtained. This was not the case for the original DREAM calorimeter \cite{Akc05a}, where the two types of fibers essentially sampled the showers in the same way. Figure \ref{rd52em2} shows that the energy resolution of the combined  signal deviates slightly from $E^{-1/2}$ scaling. The straight line fit through the data points suggests a constant term of less than 1\% \cite{Akc14b}. In any case, the energy resolution is substantially better than for either of the two individual signals, over the entire energy range covered by these measurements.
It is also better than the performance reported for other integrated em+hadronic fiber calorimeters, such as SPACAL and DREAM.
Careful analysis of the measured data showed that the contribution of sampling fluctuations to the total signal was $8.9\%/\sqrt{E}$ and that fluctuations in the number of \v{C}erenkov photoelectrons (33/GeV) increased the total stochastic term to $13.9\%/\sqrt{E}$. The small deviation from $E^{-1/2}$ scaling is due to the dependence of the scintillation response on the impact point \cite{Akc14b}. 

This impact point dependence was of no consequence for the linearity of the calorimeter response. With the exception of the lowest energy point (6 GeV, less than the minimum energy for which the beam line used for these studies was designed), the average signals were measured to be proportional to the electron energy to within $\sim 1\%$, regardless of the angle of incidence of the electrons. This is illustrated in Figure \ref{linearity_em} \cite{Akc14b}.

\begin{figure*}[htb]
\epsfysize=9.5cm
\centerline{\epsffile{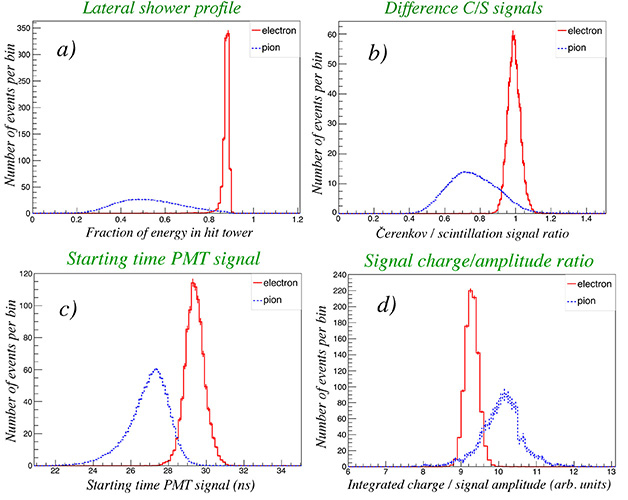}}
\caption{\small Effects of four different shower characteristics that may be used to distinguish between electron and hadron showers in the
longitudinally unsegmented RD52 fiber calorimeter. Shown are the fraction of the total signal recorded by the tower in which the particle entered ($a$),  the ratio of the \v{C}erenkov and scintillation signals of the event ($b$), the starting time of the signal in the PMT, measured with respect to an upstream trigger signal ($c$), and the ratio of the total integrated charge and the amplitude of the signal ($d$). Data obtained with 60 GeV particle beams \cite{Akc14a}.}
\label{rd52pid}
\end{figure*}

\subsubsection{\sl Particle identification.}

Traditionally, the calorimeter systems in high-energy physics experiments are separated into (at least) two sections: the electromagnetic (em) 
and the hadronic section. This arrangement offers a certain number of advantages, especially for the identification of electrons and photons,
which deposit all their energy in the em section and can thus be identified as such based on this characteristic. 

The RD52 fiber calorimeter is longitudinally unsegmented, it does not consist of separate electromagnetic and hadronic sections.
It is calibrated with electrons, and the calibration constants established in this way also
provide the correct energy for hadronic showers developing in it. This eliminates one of the main disadvantages of longitudinal segmentation, \ie the problems associated with the intercalibration of the signals from different longitudinal sections \cite{Liv17,Gan98}.
Another advantage derives from the fact that there is no need to transport signals from the upstream part of the calorimeter to the outside world. This allows for a much more homogeneous and hermetic detector structure in a $4\pi$ experiment, with fewer ``dead areas.''

Despite the absence of longitudinal segmentation, the signals provided by the RD52 fiber calorimeter offer several excellent possibilities to distinguish between different types of particles, and especially between electrons and hadrons.
Identification of isolated electrons, pions and muons would be of particular importance for the study of the decay of Higgs bosons into pairs of $\tau$ leptons, if a calorimeter of this type were to be used in an experiment at a future Higgs factory. 
\begin{figure*}[htbp]
\epsfysize=9cm
\centerline{\epsffile{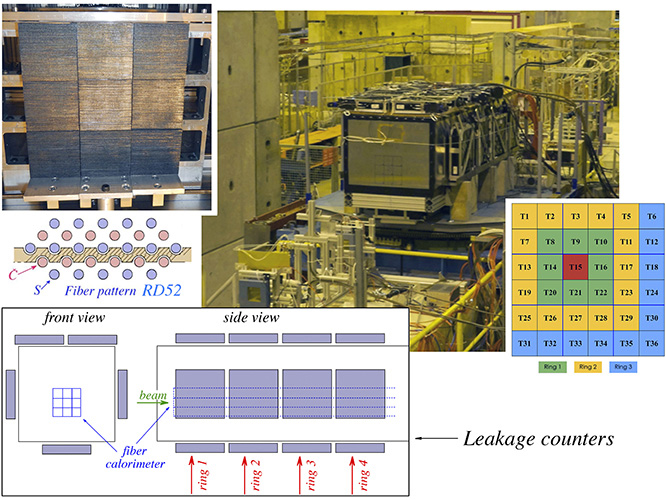}}
\caption{\small The RD52 fiber calorimeter, installed in the H8C beam area at CERN \cite{Lee17}. The system of trigger counters and beam defining elements is visible in the left bottom part of the figure. The calorimeter is surrounded on four sides by ``leakage counters,''  the layout of which is shown in the bottom left insert. The other inserts show the front face of the (lead-)fiber calorimeter (top left) and the tower structure of the readout (bottom right).}
\label{setup_CERN}
\end{figure*}

Figure \ref{rd52pid} illustrates the effects of the different identification methods \cite{Akc14a}:
\begin{enumerate}
\item
There are large differences in {\sl lateral} shower size, which can be used to distinguish between em and hadron showers.
One advantage of the RD52 calorimeter structure is that the lateral granularity can be made arbitrarily small, one can make the tower size
(defined by the number of fibers connected to one readout element) as large or small as desired. Figure \ref{rd52pid}a shows the distributions of the fraction of the shower energy deposited in a RD52 tower located on the shower axis, for 60 GeV electrons and pions.
\item The fact that both scintillation and \v{C}erenkov signals are obtained for the same showers offers opportunities to distinguish between
em and hadronic showers. For example, the ratio between the two signals is 1.0 for electrons (which are used to calibrate the signals!) and smaller than 1.0 for hadrons, to an extent determined by $f_{\rm em}$ and $e/h$. Figure \ref{rd52pid}b shows the distributions of the \v{C}erenkov/scintillation signal ratio for 60 GeV electrons and pions.
\item The next two methods are based on the fact that the light produced in the fibers travels at a lower speed ($c/n$) than the particles responsible for the production of that light, which typically travel at $c$ (see Section VII.E.2). As a result, the deeper inside the calorimeter the light is produced, the earlier it arrives at the PMT. Since the light from hadron showers is typically produced much deeper inside the calorimeter, the PMT signals start earlier than for em showers, which all produce light close to the front face of the calorimeter. Figure \ref{rd52pid}c shows distributions of the starting time of the PMT signals for 60 GeV electron and pion showers.
\item The same phenomenon also leads to a larger width of the hadron signals, since the light is produced over a much larger region in depth than for electrons. Therefore, the ratio of the integrated charge and the signal amplitude is typically larger for hadron showers. Figure \ref{rd52pid}d shows distributions of that ratio for showers induced by 60 GeV electrons and pions.
\end{enumerate}

One may wonder to what extent the different methods mentioned above are correlated, in other words to what extent the
mis-identified particles are either the same or different ones for each method. 
It turned out that by combining different $e/\pi$ separation methods, important improvements could be achieved in the capability of the longitudinally unsegmented calorimeter to identify electrons, combined with minimal contamination of mis-identified particles. 
A multivariate neural network analysis showed that the best $e/\pi$ separation achievable with the variables used for the 60 GeV beams was 99.8\% electron identification with 0.2\% pion misidentification. Further improvements may be expected by including the complete time structure information of the pulses, especially if the upstream ends of the fibers are made reflective \cite{Aco91a}.

The longitudinally unsegmented RD52 fiber calorimeter can thus be used to identify electrons with a
very high degree of accuracy.
Elimination of longitudinal segmentation offers the possibility to make a finer lateral segmentation with the same number of electronic readout channels. This has many potential benefits. A fine lateral segmentation is crucial for recognizing closely spaced particles as separate entities. Because of the extremely collimated nature of em showers (Section VII.E.1), it is also a crucial tool for recognizing electrons in the vicinity
of other showering particles, as well as for the identification of electrons in general. Unlike the vast majority of other calorimeter structures used in practice, the RD52 fiber calorimeter offers almost limitless possibilities for lateral segmentation. If so desired, one could read out every individual fiber separately. Modern silicon PM technology certainly makes that a realistic possibility (Section VIII).

\subsubsection{\sl Hadronic performance.}

The hadronic performance of the RD52 fiber calorimeter has until now only been measured with a detector that, just as its DREAM predecessor, was too small to fully contain hadronic showers. Moreover, because of problems encountered with the large-scale production of 
the required copper absorber structure, only data obtained with a 1.5 ton lead module are available at this time. The (9-module) calorimeter was subdivided into $9\times 4 = 36$ towers, and thus produced 72 signals for each event.
In order to get a handle on the shower leakage, the detector was surrounded by an array of 20 plastic scintillation counters (measuring 50$\times$50$\times$10 cm$^3$ each).
Figure \ref{setup_CERN} shows a picture of the setup in which this detector combination was tested at CERN.
\begin{figure*}[htbp]
\epsfysize=8cm
\centerline{\epsffile{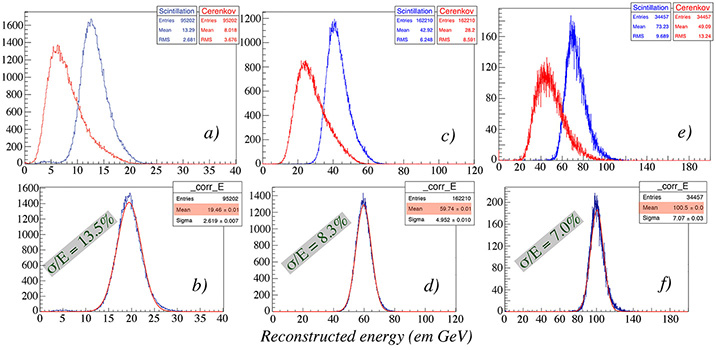}}
\caption{\small
Signal distributions for $\pi^-$ beam particles of 20, 60 and 100 GeV showering in the RD52 lead-fiber calorimeter. The top row ($a,c,e$) shows the signal distributions measured for the scintillation (S) and \v{C}erenkov (C) signals. The S signals are, on average, larger, and their distribution is less asymmetric.The bottom row ($b,d,f$) shows the signal distributions that were obtained after combining the S and C distributions according to Equation \ref{eq5}, with $\chi = 0.45$ \cite{DRE13}.}
\label{rd52had}
\end{figure*}

As usual, all 72 calorimeter signals were calibrated with electrons. The leakage counters were calibrated with a muon beam, the muons deposited on average 100 MeV in each module they traversed. Next, hadron beams were sent into the central region of the calorimeter. Figure \ref{rd52had} shows the scintillation and \v{C}erenkov signal distributions for 20, 60 and 100 GeV $\pi^-$ showers, as well as the ones derived on the basis of the measured em shower fraction, using Equation \ref{eq5} \cite{DRE13}. The latter distributions exhibit the familiar benefits of the dual-readout method: a relatively narrow, Gaussian signal distribution centered around the correct mean value, \ie the energy of the electrons that were used to calibrate the channels. The energy resolution is not very different from the one obtained with the original DREAM calorimeter
(Figure \ref{QSafter}), which is no surprise since in both cases leakage fluctuations were a dominant contribution to the hadronic energy resolution.
\begin{figure}[b!]
\epsfysize=7cm
\centerline{\epsffile{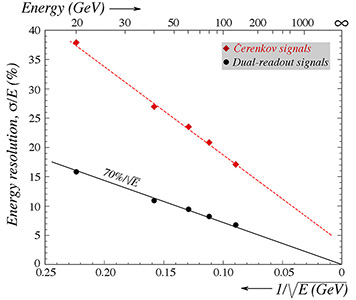}}
\caption{\small The hadronic energy resolution of the RD52 lead-fiber dual-readout calorimeter, for single pions. Shown are the results for the \v{C}erenkov signals alone, and for the dual-readout signals, obtained with Eq. \ref{eq5} \cite{Lee17}.}
\label{reshad}
\end{figure}

Figure \ref{reshad} shows that the hadronic energy resolution obtained with the dual-readout method scales with $E^{-1/2}$, unlike the resolution obtained with the individual signals. The same phenomena were observed with the DREAM calorimeter (Figure \ref{QSafter}c). The data analysis showed that the resolution improved significantly 
when the signals from the leakage counters were taken into account \cite{Lee17}, despite the fact that these counters only provided a very crude and incomplete measurement of the shower energy leaking out of the fiber structure (see Figure \ref{setup_CERN}). This illustrates that leakage fluctuations were indeed a dominant contribution to the results shown in Figure \ref{reshad}.

In order to estimate the improvement that may be expected in a calorimeter that is large enough to contain the showers at the required level, elaborate GEANT4 based Monte Carlo simulations have been performed. The reliability of these simulations was assessed by comparing the results with the experimental data obtained with the DREAM calorimeter\cite{Akc14c}.
It turned out that the \v{C}erenkov response function (Figure \ref{QSbefore}b) was well described by these simulations. On the other hand, the simulated scintillation distribution was more narrow, less asymmetric and peaked at a lower value than for the experimental data (Figure \ref{QSbefore}a). This is believed to be due to the fact that the
non-relativistic component of the shower development, which is completely dominated by processes at the nuclear level, is rather poorly described by GEANT4, at least by the standard FTFP\_BERT hadronic shower development package. Both the average size of this component, as well as its event-to-event fluctuations, are at variance with the experimental data. This non-relativistic shower component only plays a role for the scintillation signals, {\sl not} for the \v{C}erenkov ones.

Yet, some aspects of hadronic shower development that are important for the dual-readout application were found to be in good agreement with the experimental data, \eg the shape of the \v{C}erenkov response function and the radial shower profiles. Attempts to use the dual-readout technique on simulated shower data reasonably reproduced some of the essential characteristics and advantages of this method: a Gaussian response function, hadronic signal linearity and improved hadronic energy resolution. The fact that the reconstructed beam energy was systematically too low may be ascribed to the problems with the non-relativistic shower component mentioned above.
As stated above, the main purpose of these very time consuming simulations was to see if and to what extent the hadronic performance would improve as the detector size is increased. Figure \ref{ftfp}a shows the signal distribution obtained for 100 GeV $\pi^-$ in a copper-based RD52 calorimeter with a lateral cross section of  65$\times$65 cm$^2$. The mass of such a (10$\lambda_{\rm int}$ deep) device would be  $\sim 6$ tonnes. According to these simulations, the average calorimeter signal,
reconstructed with the dual-readout method, would be 90.2 GeV, and the energy resolution would be 4.6\%. 
\begin{figure}[htb]
\epsfysize=11cm
\centerline{\epsffile{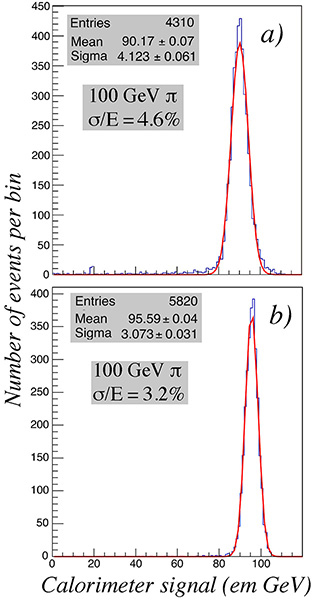}}
\caption{\small GEANT4 simulations of the response function to 100 GeV $\pi^-$ particles of a dual-readout fiber calorimeter with the 
RD52 structure, and lateral dimensions of 65$\times$65 cm$^2$ \cite{Akc14c}. Results are shown for the standard FTFP\_BERT hadronic shower simulation package ($a$), and with the high-precision version of this package, FTFP\_BERT\_HP ($b$). From reference \cite{Wig16}.}
\label{ftfp}
\end{figure}

In order to see to what extent these simulations depend on the choice of the hadronic shower development package,  the simulations were repeated using the high precision version of the hadronic shower simulation package (FTFP\_BERT\_HP), which seems to provide a much more elaborate treatment of the numerous neutrons produced in the shower process. Indeed, the results of this work (Figure \ref{ftfp}b) show a clear improvement: the average calorimeter signal increased to 95.6 GeV,
and is thus within a few percent equal to that of an em shower developing in the same calorimeter structure (one of the crucial advantages of calorimeters based on the DREAM principle). Also the energy resolution improved significantly, from 4.6\% to 3.2\%.
Simulations for 200 GeV hadron showers with the FTFP\_BERT\_HP package yielded an average signal of 191 GeV and an energy resolution of 2.4\% \cite{Akc14c,Wig16}. 
\begin{figure*}[htbp]
\epsfysize=9.5cm
\centerline{\epsffile{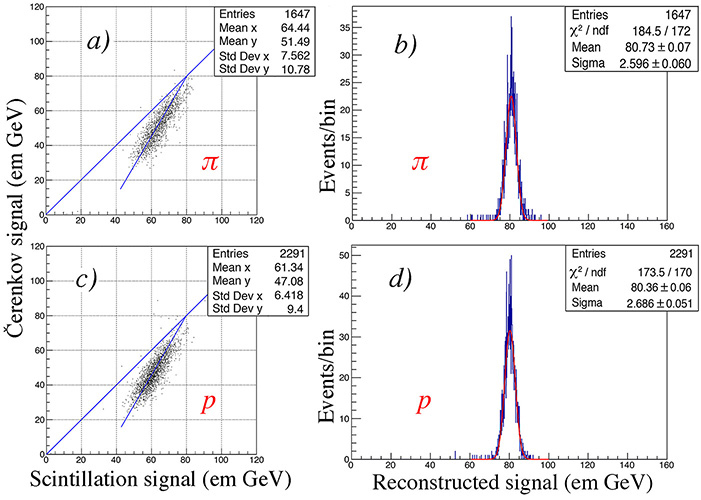}}
\caption{\small{Scatter plots of the \v{C}erenkov \vs the scintillation signals from showers induced by 80 GeV $\pi^+$ ($a$) and 80 GeV protons ($c$) in the RD52 lead-fiber calorimeter. Projection of the rotated scatter plots on the $x$ axis for the pions ($b$) and protons ($d$). The rotation procedure was identical to that used for 60 GeV $\pi^-$ (Figure \ref{DRrotation}). Experimental data from \cite{Lee17}}}
\label{80ppi}
\end{figure*}

These simulations thus suggest that resolutions of a few percent are indeed feasible, and that the hadronic performance of a sufficiently large copper-based RD52 calorimeter would be at the same level as that of the compensating SPACAL and ZEUS calorimeters, or even better.
It should also be emphasized that the results shown in Figure \ref{ftfp} are for {\sl single hadrons}. 
As explained in Section III, the {\em jet} energy resolution of copper-based dual-readout fiber calorimeters may also be expected to be much better than that of high-$Z$ compensating calorimeters \cite{Wig13}, since the difference between the calorimeter responses to showers and mips is much smaller.
For the copper-based DREAM calorimeter, an $e/mip$ value of 0.84 was measured \cite{Akc04}, \vs 0.62 and 0.72 for the compensating ZEUS \cite{Dre90} and SPACAL \cite{Aco92c}
calorimeters, respectively.
We recall that the possibility to measure jets with superior resolution compared to previously built high-$Z$ compensating calorimeters was one of the main reasons why the dual-readout project was started.

\subsubsection{\sl Results obtained with the rotation method}

In this subsection, some results are shown that were obtained by the RD52 Collaboration with the rotation method described in Section V, and graphically illustrated in Figure \ref{dr5}.
This method can be used for an ensemble of mono-energetic hadron events, as typically available in beam tests of calorimeter modules. There is no need to know the energy of these hadrons, since this follows from the intersection of the line around which the hadronic data points are clustered in the $S - C$ scatterplot and the line $C = S$, where all electron events are located. Rotation of the hadronic data around this point ($P$) by a fixed (energy independent) angle leads to a very narrow, Gaussian signal distribution centered around the correct energy value.
Figure \ref{DRrotation} shows an example of the results of this procedure, for 60 GeV $\pi^-$. In the following, some other results are shown \cite{Lee17}.
\begin{figure*}[htbp]
\epsfysize=9.5cm
\centerline{\epsffile{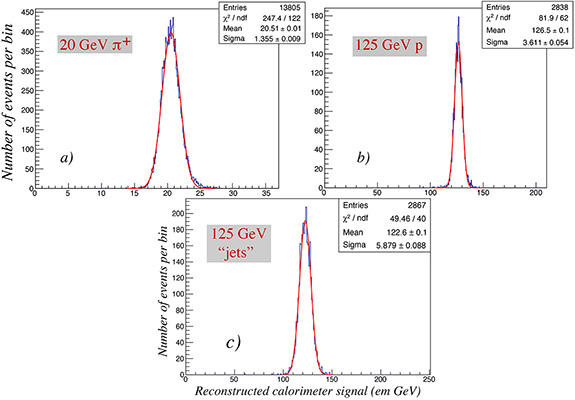}}
\caption{\small{Signal distributions from the RD52 lead-fiber calorimeter for 20 GeV $\pi^+$ ($a$), 125 GeV protons ($b$) and 125 GeV multiparticle events ($c$) obtained with the rotation method described in the text. The energy scale is set by electrons showering in this detector \cite{Lee17}.}}
\label{20pi125p+j}
\end{figure*}

Figure \ref{80ppi} shows the \v{C}erenkov \vs scintillation scatter plots for the 80 GeV $\pi^+$ (Figure \ref{80ppi}a) and proton (Figure \ref{80ppi}c) data. These plots show a significant difference between the pion and proton signal distributions. The average \v{C}erenkov signal is about 10\% larger for the pions than for the protons, a consequence of the absence of leading $\pi^0$s in the proton showers \cite{Akc97}.
However, using the intersection of the axis of the locus of the events in the scatter plot and the $C/S = 1$ point as the center of rotation, and the same rotation angle ($\theta$) as for 60 GeV, the resulting signal distributions turned out to have about the same average value: 80.7 GeV for the pions (Figure \ref{80ppi}b) and 80.4 GeV for the protons (Figure \ref{80ppi}d). The widths of both distributions were also about the same: 2.60 GeV for pions, 2.69 GeV for protons.
Regardless of the differences between the production of $\pi^0$s (and thus of \v{C}erenkov light) in these two types of showers,
the signal distributions obtained with the dual-readout procedure were thus practically indistinguishable.
This feature is in stark contrast with results obtained with other types of (non-compensating) calorimeters. 
For example, ATLAS has reported significant differences between the calorimeter response functions for high-energy pions and protons \cite{Adr09}. Whereas the response was systematically larger for the pions (2 - 5\%, between 50 and 180 GeV), the energy resolution was 
significantly better for the protons. Even larger differences were reported for prototype studies of the CMS Very Forward Calorimeter \cite{Akc98}. 

The rotation method was also applied for 20 GeV, 40 GeV and 125 GeV particles, with very similar results.
Also here, the average \v{C}erenkov signals in the raw data were significantly smaller for protons than for pions.
However, after applying the same rotation procedure as for the 60 and 80 GeV data (always using the same rotation angle ($\theta$), the resulting signal distributions were centered around approximately the correct values. The fact that the rotation angle used to achieve these results is independent of the particle type and the energy is consistent with Groom's observation that this angle only depends on the energy independent value of the $\chi$ parameter defined in Equation \ref{eq4} \cite{PDG16}.

The same method was also used for multiparticle events, samples of which were available for beam energies of +40, +60, +100 and +125 GeV.
During these dedicated runs, the beam hadrons were required to produce a signal of at least 6 mip in the scintillation counter downstream of the interaction target, while producing a mip signal directly upstream of this target. No distinction was made between protons and pions for this analysis. Otherwise, the conditions were identical to the ones used for the single-hadron analysis.
\begin{figure}[htbp]
\epsfysize=11.5cm
\centerline{\epsffile{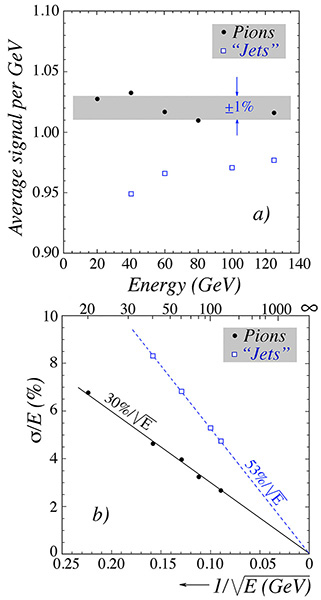}}
\caption{\small{The average calorimeter signal per GeV ($a$) and the fractional width of the signal distribution ($b$) as a function of energy, for single pions and multiparticle events (``jets''). Results are given for the RD52 dual-readout calorimeter signals, obtained with the rotation method \cite{Lee17}.}}
\label{jetresults}
\end{figure}

Figure \ref{20pi125p+j} shows the dual-readout signal distributions measured for 20 GeV $\pi^+$, 125 GeV protons and 125 GeV multiparticle events. 
The results exhibit the following features, which are illustrated in Figure \ref{jetresults}:
\begin{itemize}
\item The calorimeter is very linear, both for pion and proton detection. The beam energy is correctly reconstructed at all energies within a few percent, using the energy scale for electrons, \ie the particles that were used to calibrate the signals.  The vertical scale is normalized to the electron response. The hadron signals are thus a 
few percent larger than those for em showers of the same energy. 
\item The reconstructed energies are somewhat lower in the case of the multiparticle events, more so at low energy (Figure \ref{jetresults}a). Very substantial differences with the single hadron results are observed in the size of the \v{C}erenkov component, which is on average considerably smaller for the multiparticle events. 
\item The reconstructed signal distributions are very narrow, narrower  than those reported by any other detector we know of.
\item The reconstructed signal distributions are very well described by Gaussian functions.The normalized $\chi^2$ values
varied between 1.02 and 2.27 for all particles and ``jets''.
\item The fractional width of the reconstructed signal distribution also scales very well as expected for an energy resolution dominated by Poisson fluctuations. Over the full energy range of 20 -- 125 GeV, $\sigma/E$ was measured to be $(30 \pm 2\%)/\sqrt{E}$ for single pions and protons, and $53\%/\sqrt{E}$ for ``jets''. 
\end{itemize}

The differences between the results for single hadrons and for multiparticle events can be understood by realizing that the primary interaction of the beam particles in the case of the multiparticle events took place at a distance of about 75 cm upstream of the calorimeter.
Low-energy secondaries produced in these interactions may have traveled at such large angles with the beam line that they physically missed the calorimeter, as well as the leakage counters surrounding the calorimeter. The effect of that is larger when the energy of the incoming beam particle is smaller. The increased side leakage is probably also the main factor responsible for the increased width of the signal distribution.
The difference in the relative strength of the \v{C}erenkov component most likely reflects the fact that the average energy fraction carried  by the em component in hadronic showers increases with energy. Therefore, if the energy of the incoming beam particle is split between at least six secondaries (which was the trigger condition for multiparticle events), the total em energy fraction is likely to be smaller than when the beam particle enters the calorimeter and deposits its entire energy there in the form of a single hadronic shower.  

\subsection{Other RD52 results}

Detailed studies with the fine-grained RD52 fiber calorimeter have revealed important information about the showering particles that are of interest for other calorimeters as well. In this subsection, the em shower profiles and the time structure of the showers are addressed.

\subsubsection{\sl The electromagnetic shower profiles.}

The fine-grained RD52 fiber calorimeters lend themselves very well for precision measurements of the lateral shower profiles. 
\begin{figure}[htbp]
\epsfysize=6cm
\centerline{\epsffile{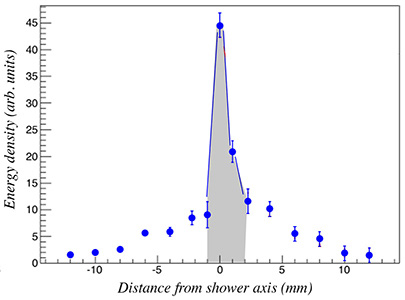}}
\caption{\small
The lateral profile of 100 GeV electron showers in the RD52 lead-fiber calorimeter, measured with the scintillation signals \cite{Akc14b}.}
\label{emprofilerd52}
\end{figure}
This was done by 
moving a 1 mm wide electron beam across the boundary between neighboring towers and measuring the energy fraction deposited in each of these towers. This narrow beam was obtained by selecting beam particles based on the coordinates of the points where they traversed upstream wire chambers. Figure \ref{emprofilerd52} shows the profile measured for 100 GeV electrons in the lead-based RD52 calorimeter.
Since the calorimeter is longitudinally unsegmented, the profile is integrated over the full depth. It exhibits a very pronounced central core, which is presumably caused by the extremely collimated nature of the showers in the early stage of the shower development, before the shower maximum is reached. In this stage, the shower mainly consists of energetic bremsstrahlung photons, which convert into energetic $e^+e^-$ pairs that travel in the same direction as the beam particles. According to Figure \ref{emprofilerd52}, a considerable fraction of the shower energy ($\sim 20\%$) is deposited in a cylinder with a radius of 1 mm about the shower axis \cite{Akc14b}.

This feature has important consequences for this type of calorimeter, where the distance between neighboring fibers of the same type is 2-3 mm (see Figure \ref{dreamstructure}). The calorimeter signal (from this early shower component) depends crucially on the impact point of the beam particles, if these enter the calorimeter parallel to the fibers. This dependence is quickly reduced when the electrons enter the calorimeter at a small angle with the fibers. As the angle increases, this early collimated shower component is sampled more and more in the same way as the rest of the shower. However, at angles where this is not the case, this effect adds an additional component to the em energy resolution. This is clearly observed in Figure \ref{angdep}, which shows the energy resolution for 20 GeV electrons as a function of the angle of incidence \cite{Car16}.
This effect is, in first approximation, energy independent and thus results in a constant term in the em energy resolution. The measured em energy resolution of the scintillation signals of the RD52 copper-fiber calorimeter (Figure \ref{rd52em2}) exhibits indeed a clear deviation from $E^{-1/2}$ scaling.
Because of the extreme dependence on the angle of incidence, one should be careful when comparing the em performance measured with different fiber calorimeters. For example, the improvement in the em scintillation resolution of the RD52 calorimeter with respect to the DREAM one is much larger than suggested by the comparison in Figure \ref{rd52em1}, because the angles at which the DREAM measurements were performed were twice as large as in case of the RD52 calorimeter. The distance separating neighboring fiber clusters in DREAM was such that the position dependence of the scintillation signal in these measurements even led to a non-Gaussian response function (Figure \ref{rd52em1}c).
\begin{figure}[b!]
\epsfysize=8cm
\centerline{\epsffile{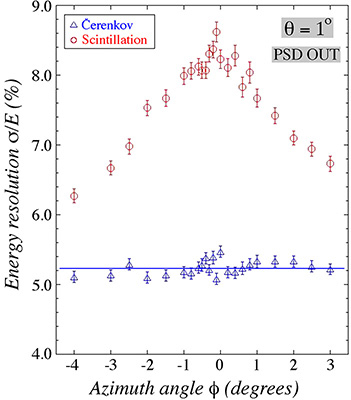}}
\caption{\small The energy resolution measured for 20 GeV electrons in the scintillation and the \v{C}erenkov channels of the RD52 copper-fiber calorimeter, as a function of the azimuth angle of incidence ($\phi$) of the beam particles. The tilt angle $\theta$ was $1^\circ$ \cite{Car16}.}
\label{angdep}
\end{figure}

Now, why does this position dependence of the response function only affect the resolution measured with the scintillation signals? The reason is that the collimated early shower component does {\sl not} contribute to the \v{C}erenkov signals, since the \v{C}erenkov light produced by shower particles traveling in the same direction as the fibers falls outside the numerical aperture of the fibers. For the beam electrons, the \v{C}erenkov fibers thus only registered shower particles that traveled at relatively large angles with the shower axis ($20 - 60^\circ$), and such particles are for all practical purposes almost exclusively found beyond the shower maximum, where the shower is wide compared to the typical distance separating neighboring fibers of the same type. The ``constant'' term that affects the scintillation resolution is thus practically absent for the \v{C}erenkov signals, as illustrated by Figure \ref{rd52em2}.
The different effects of the angle of incidence on the two types of calorimeter signals is made very clear in Figure \ref{angdep}, which shows the energy resolution for 20 GeV electron showers as a function of the angle of incidence of the beam particles.
Whereas the resolution measured with the \v{C}erenkov signals is independent of that angle, the resolution measured with the scintillation signals increases dramatically when the particles enter the calorimeter in approximately the same direction as the fiber
axes.
\begin{figure}[htb]
\epsfysize=10.5cm
\centerline{\epsffile{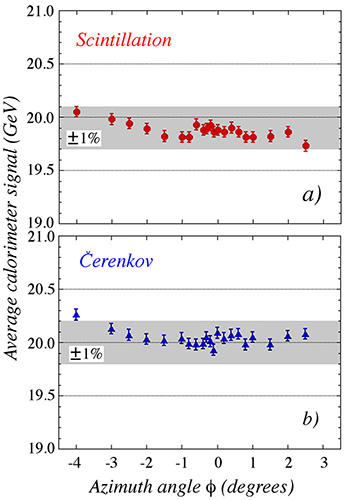}}
\caption{\small The average signals for 20 GeV electrons in the scintillation ($a$) and the \v{C}erenkov ($b$) channels of the RD52 copper-fiber calorimeter, as a function of the azimuth angle of incidence ($\phi$) of the beam particles. The tilt angle $\theta$ was $1^\circ$, and the PSD was in the beam line. The shaded area represents a variation of $\pm 1\%$ with respect to the average signal \cite{Car16}.}
\label{emresponse}
\end{figure}
On the other hand, the calorimeter response is not sensitive to the angle of incidence. Measurements of the average signal per GeV deposited energy for 20 GeV electrons showed it to be the same to within 1\% for angles ranging from $-4^\circ$ to $4^\circ$, both for the scintillation and the \v{C}erenkov signals (Figure \ref{emresponse}). This indicates that the small position dependent differences in sampling fraction for the early shower component, which affect the energy resolution, do not translate into systematic response differences as a function of the angle of incidence.

In Figure \ref{DREAMmu}, another consequence of this difference between the two types of signals from dual-readout calorimeters is shown.
When muons traverse this calorimeter parallel to the fibers, the \v{C}erenkov fibers only register the radiative component of their energy loss, because the \v{C}erenkov light emitted in the non-radiative (ionizing) component falls outside the numerical aperture of the fibers.

These results also have consequences for other types of calorimeters. Typically, the lateral granularity is chosen on the basis of the Moli\`ere radius of the calorimeter structure, with the argument that this parameter determines the radial shower development. However, the results shown here indicate that em showers have a very pronounced, extremely collimated core, whose radial dimensions are very small compared to the Moli\`ere radius. A much finer granularity would make it possible to resolve doublets, or recognize electromagnetic components of jets much better than in a calorimeter with a granularity based on the value of the Moli\`ere radius. Fibers  offer this possibility, since the lateral granularity of a calorimeter of the RD52 type could be made arbitrarily small. 

\subsubsection{\sl Time structure of the showers.}

Earlier in this section, it was shown how the time structure of the calorimeter signals could provide crucial information. It could, for example, be used 
\begin{itemize}
\item to distinguish between and separate the scintillation and \v{C}erenkov components of the light signals from crystals (Figures \ref{Modoping}, \ref{BGOtime}),
\item to identify showers initiated by electrons and photons (Figure \ref{rd52pid}), and
\item to recognize and measure the contribution of neutrons to the calorimeter signals (Figure \ref{SQ11}).
\end{itemize}
The timing information is particularly important for fiber calorimeters such as the ones discussed here.
Even though light attenuation is not a big effect in the optical fibers used as active media in these calorimeters, it may have significant consequences for the hadronic performance. The depth at which the light is produced in these showers fluctuates at the level of a nuclear interaction length, \ie effectively $\sim$25 cm in the RD52 fiber calorimeters. The light attenuation length in the scintillating fibers amounts to $\sim 8$ m, while in the \v{C}erenkov ones values up to 20 m have been measured. But even for an attenuation length of 20 m, the mentioned depth fluctuations introduce a constant term of $\sim 1\%$ in the hadronic energy resolution, and this term increases correspondingly for shorter attenuation lengths. If one could measure the depth at which the light is produced, the signals could be corrected event by event for the effects of light attenuation. 
The timing information of the calorimeter signals provides this information, thanks to the fact that light in the optical fibers travels at a lower speed than the particles that generate this light.
\begin{figure}[htb]
\epsfxsize=8.5cm
\centerline{\epsffile{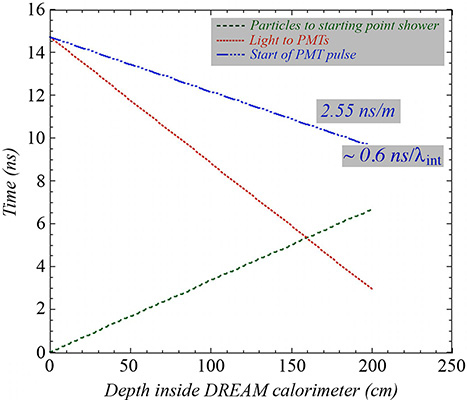}}
\caption{\small Dependence of the starting time of the PMT signals on the average depth ($z$) inside the calorimeter where the light is produced (the dash-dotted line). This time is measured with respect to the moment the particles entered the calorimeter. Also shown are the time it takes the particles to travel to $z$ (the dashed line) and the time it takes the light to travel from $z$ to the PMT (the dotted line) \cite{Akc14a}.}
\label{timing1}
\end{figure}
The effective speed of light generated in the fibers is $c/n$, with $n$ the index of refraction. For an index of 1.59, typical for polystyrene-based fibers, this translates into a speed of 17 cm/ns. On the other hand, the shower particles responsible for the generation of light in the fibers typically travel at a speed close to $c$. The effects of this are illustrated in Figure \ref{timing1}, which shows how the starting time of the PMT signal varies with the (average) depth at which the light is produced inside the calorimeter. 
The deeper inside the calorimeter the light is produced, the earlier the PMT signal starts. For the polystyrene fibers, this effect amounts to 2.55 ns/m. For the RD52 lead-fiber calorimeter, this corresponds to $\sim 0.6$ ns/$\lambda_{\rm int}$.
\begin{figure*}[htb]
\epsfysize=8.5cm
\centerline{\epsffile{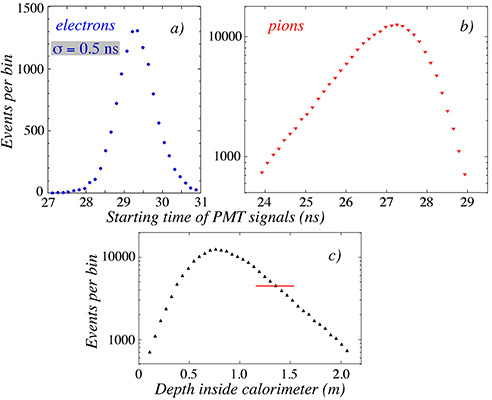}}
\caption{\small The measured distribution of the starting time of the DREAM calorimeter's scintillation signals produced by 60 GeV electrons ($a$) and 60 GeV pions ($b$). This time is measured with respect to the moment the beam particle traversed trigger counter T1, installed upstream of the calorimeter. These data were used to determine the distribution of the average depth at which the light was produced in the hadron showers ($c$). The horizontal red line is an error bar that represents the precision with which the (average) depth of the light production in an individual event can be determined \cite{Akc14a}.}
\label{timing2}
\end{figure*}

This was experimentally verified with 60 GeV electron and pion event samples, using a Time-to-Digital Converter (TDC) \cite{Akc14a}. The TDC was started by the signal produced by an upstream trigger, and stopped by the signal from the central calorimeter tower. Figure \ref{timing2}a shows the TDC signal distribution for the electron showers. In these showers, the light was, on average, produced at a depth of $\sim$12 cm inside the calorimeter ($10 X_0$),
with event-to-event variations at the level of a few cm. The width of this distribution, $\sim$0.5 ns, is thus a good measure for the precision with which the depth of the light production can be determined for individual events, $\sim$20 cm.

Figure \ref{timing2}b shows the measured TDC distribution for 60 GeV $\pi^-$. This distribution peaked $\sim$1.5 ns earlier than that of the electrons, which means that the light was, on average, produced 60 cm deeper inside the calorimeter. The distribution is also asymmetric, it
has an exponential tail towards early starting times, \ie light production deep inside the calorimeter. This measured TDC signal distribution could be used to reconstruct the average depth at which the light was produced for individual pion showers. The result, shown in Figure \ref{timing2}c, essentially represents the longitudinal shower profile of the 60 GeV pion showers in this calorimeter \cite{Akc14a}.

In earlier studies of longitudinally unsegmented calorimeters, the depth of the light production was measured from
the displacement of the lateral center-of-gravity of the shower with respect to the entrance point of the beam particles. To use this method, it was necessary to rotate the calorimeter over a small angle with respect to the beam line \cite{Akc05b}. The study described here does not require such a rotation. And unlike the displacement method, it also works for jets and neutral particles.
\begin{figure}[htbp]
\epsfysize=9cm
\centerline{\epsffile{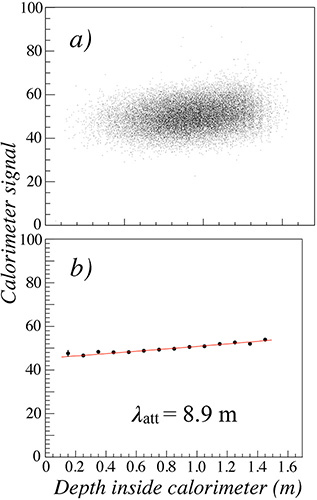}}
\caption{\small Light attenuation in the fibers. The scatter plot ($a$) shows the calorimeter signal for \v{C}erenkov light from 80 GeV $\pi^-$ versus the average depth at which that light was produced inside the DREAM calorimeter. This depth was determined from the timing information. The average signal, as a function of depth, provides the effective light attenuation curve of the fibers ($b$) \cite{Akc14a}.}
\label{latt}
\end{figure}

Figure \ref{latt} shows results of measurements performed to assess the effects of light attenuation in the fibers. The scatter plot in Figure \ref{latt}a represents the calorimeter signal for the \v{C}erenkov light from 80 GeV $\pi^-$ versus the average depth at which that light was produced inside the calorimeter. As the light is produced deeper inside, the signal tends to be, on average, somewhat larger.  This effect is quantified in Figure \ref{latt}b, which shows the average signal as a function of the depth at which the light was produced. The data points are well described with an exponential curve with a slope of 8.9 m, which thus represents the (effective) attenuation length of these fibers. This means that the signal changes by 2-3\%/$\lambda_{\rm int}$ as a result of light attenuation. And since this calorimeter is intended for hadronic energy measurements at the level of 1\%, elimination of the energy independent term caused
by light attenuation effects is important.
\vskip 2mm

Until a few years ago, detailed measurements of the time structure of the calorimeter signals required a high-quality digital sampling oscilloscope \footnote{The results shown in Figures \ref{Modoping}, \ref{BGOtime} and \ref{SQ11} were all performed with a 
Tektronix TDS 7254B digital oscilloscope, which provided a sampling capability of 5 GSample/s, at an analog bandwidth of 2.5 GHz, \ie the signals were sampled every 0.4 ns. }.
In recent years, developments in microelectronics have made it possible to obtain this type of capability for a fraction of the cost.
For example, CAEN is now offering a 36-channel VME module (V1742),  based on the DRS4 chip \cite{Rit10}, which provides 5 GSample/s sampling. 
The RD52 Collaboration used such a module to measure the time structure of 30 different calorimeter signals simultaneously \cite{Wig16}.

\begin{figure}[htb]
\epsfysize=8.2cm
\centerline{\epsffile{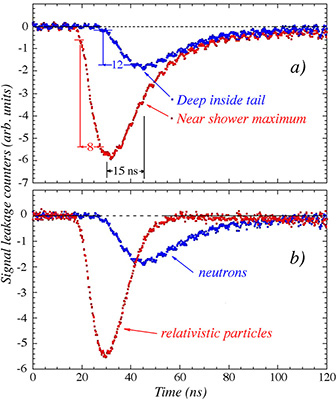}}
\caption{\small Average time structure of the signals measured in leakage counters surrounding the RD52 lead-fiber calorimeter, for 40 GeV $\pi^-$ steered into the center of this calorimeter.
Diagram $a$ shows the signals measured in a counter located close to the shower maximum (not far from the front face of the calorimeter) and in a counter located near the shower tail,
\ie about 2 m from the front face of the calorimeter. In diagram $b$, the signal from the upstream counter is unfolded into a ``neutron" and a ``prompt'' component \cite{Wig16}.}
\label{leaktime}
\end{figure}
Figure \ref{leaktime} shows one result of these measurements, which concerns the time structure of the 
average signals recorded in two different leakage counters. These counters (see Figure \ref{setup_CERN}) were located close to the shower maximum (the early signal), and near the end of the calorimeter module (the late signal). The latter signal consisted very likely exclusively of recoil protons produced by elastic neutron scattering, while the early signal may also contain a contribution from relativistic particles produced in the shower development and escaping the calorimeter. In the hadronic shower development, typically a few thousand neutrons are released from the nuclei in which they were bound. They typically carry a few MeV kinetic energy and lose that energy predominantly by means of elastic scattering off protons in the plastic components of the detectors, with a time constant of $\sim$10 ns. The time difference between the two signals shown in Figure \ref{leaktime} and the difference in rise time are consistent with the above assessment.

These are only a few examples of the information that can be obtained on the basis of time information about the calorimeter signals. 
We expect that more applications will be developed, especially if faster light detectors become available.

\subsection{Spin-off effects}

The results obtained with the DREAM and RD52 detectors have inspired a number of ideas to use the benefits offered by the dual-readout technique in alternative ways. 
The ideas that have been proposed are summarized below. They are all based on schemes with a higher \v{C}erenkov light yield than in the fiber detectors. This is achieved by making the detector fully active.
\begin{itemize}
\item Para and coworkers \cite{para} have proposed a homogeneous calorimeter made of small (few cm) cubic scintillating crystals, read out by SiPMs using UV and visible light filters. This structure allows for a fine lateral and longitudinal segmentation and is aimed at application in a PFA environment \cite{Magill}. 
In this context, it should be mentioned that Groom has demonstrated that application of the dual-readout technique in a homogeneous calorimeter leads to a degradation of the hadronic energy resolution, compared to that of a sampling detector with organic scintillator readout \cite{groompara}.
\item
Takeshita \cite{Take} has proposed a sandwich calorimeter in which plastic scintillator plates are alternated with lead glass plates. The latter serve as absorber material and produce \v{C}erenkov light as well.
\item The ADRIANO (A Dual-Readout Integrally Active Non-segmented Option) R\&D project \cite{gatto} aims for a design similar to SPACAL \cite{Aco91c} with scintillating fibers  embedded in a matrix made out of heavy glass instead of lead.
\end{itemize}
Small prototypes have been built and tested in the context of all these R\&D projects.
However, until now none of these ideas has resulted in a practical detector of which the performance can be tested. 

\section{Assessment and outlook}

\subsection{Dual-readout \vs compensation}

The dual-readout technique was developed with the goal to obtain calorimeter systems that offer the same advantages as compensating systems, without the associated disadvantages of the latter. The advantages deriving from compensation include signal linearity, Gaussian response functions and excellent energy resolution for hadron showers. The excellent linearity achieved with the RD52 fiber calorimeters is illustrated in Figure \ref{linearity_em} for electromagnetic showers and in Figure \ref{jetresults}a for hadrons, while Figure \ref{20pi125p+j} shows that the hadronic response functions obtained with this calorimeter are very well described by Gaussian functions.
\begin{figure*}[hbt]
\epsfysize=8.5cm
\centerline{\epsffile{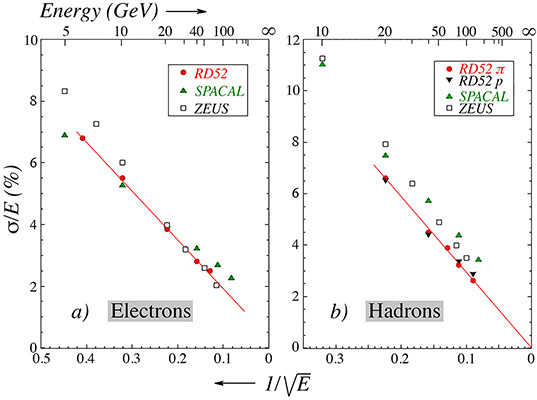}}
\caption{\small
Energy resolutions reported for the detection of electrons($a$) and hadrons ($b$) by RD52 \cite{Akc14b,Lee17}, SPACAL \cite{Aco91c} and ZEUS \cite{Beh90}. From reference \cite{LLW17}.}
\label{Erescomp}
\end{figure*}

In Figure \ref{Erescomp}, the energy resolutions obtained with the best compensating calorimeters, ZEUS \cite{Beh90} and SPACAL \cite{Aco91c}, are compared with the results obtained with the RD52 fiber calorimeter. Figure \ref{Erescomp}b shows that the hadronic RD52 values are actually better than the ones reported by ZEUS and SPACAL, while Figure \ref{Erescomp}a shows that the RD52 em energy resolution is certainly not worse.

In making this comparison, it should be kept in mind that
\begin{enumerate}
\item The em energy resolutions shown for RD52 were obtained with the calorimeter oriented at a much smaller angle with the beam line 
($\theta,\phi = 1^\circ ,1.5^\circ$) than the ones for SPACAL ($\theta,\phi = 2^\circ ,3^\circ$) \cite{Akc14b}. As shown in Figure \ref{angdep},
the em energy resolution is extremely sensitive to the angle between the beam particles and the fiber axis when this angle is very small.
\item The instrumented volume of the RD52 calorimeter (including the leakage counters) was less than 2 tons, while both SPACAL and ZEUS
obtained the reported results with detectors that were sufficiently large ($> 20$ tons) to contain the showers at the 99+\% level. As stated before, the hadronic resolutions shown for RD52 are dominated by fluctuations in lateral shower leakage, and a larger instrument of this type is 
thus very likely to further improve the results.
\end{enumerate}
The comparison of the hadronic results seems to indicate that the dual-readout approach offers even better opportunities to achieve superior hadronic performance than compensation. One may wonder why that is the case. Here is our explanation.

The main reason for the poor hadronic energy resolution of the calorimeters used in high-energy physics experiments are fluctuations in the response to the non-em shower component. These are dominated by fluctuations in the {\sl invisible energy}, \ie the energy needed to release nucleons from the atomic nuclei in which they are bound when these nuclei are subject to nuclear reactions in the shower development process. Compensating calorimeters and dual-readout calorimeters both try to eliminate/reduce the effects of these fluctuations on the signal distributions by means of a measurable variable that is correlated to the invisible energy. However, the variables used for this purpose are different in compensating and dual-readout calorimeters.

Compensating calorimeters exploit the fact that the total kinetic energy carried by the neutrons produced in the shower development
is correlated to the total invisible energy loss. Especially in high-$Z$ absorber materials, this correlation is quite strong, since
a very large fraction of the nucleons released in the nuclear reactions are neutrons in this case. First of all, the neutron-to-proton ratio
is larger in high-$Z$ nuclei (\eg 1.5 in lead, \vs 1.1 in iron and copper). Second, and more importantly, the large Coulomb barrier strongly favors neutrons
in the evaporation phase of the nuclear reactions. It is estimated that neutrons outnumber protons by a factor of 10 when high-energy hadrons are absorbed in lead \cite{Wig17,Ler86}. The correlation between the nuclear binding energy losses and the total kinetic energy carried by neutrons is supported by the results of Monte Carlo simulations, shown in Figure \ref{brau} for 5 GeV $\pi^-$ absorption in $^{238}$U, the highest-$Z$ nucleus available in nature. Interestingly, the correlation is in that case less strong than in lead, since a large fraction of the neutrons in uranium are the result of fission reactions, and have nothing to do with nuclear binding energy losses. The advantage of lead over uranium in this respect was clearly demonstrated by the ZEUS Collaboration \cite{Dre90}.

In dual-readout calorimeters, the correlation between the fraction of the hadron energy carried by em shower components 
(initiated by $\pi^0s, \eta$s, energetic $\gamma$-rays) and the nuclear binding energy losses is exploited with the goal to improve the hadronic energy resolution of the calorimeter. In this case, perfect correlation thus means that the total nuclear binding energy losses represent a fixed fraction of the non-em shower energy.

Apparently, in hadronic shower development the correlation with the total nuclear binding energy loss is stronger for the 
total non-em energy (derived from the em shower fraction) than for the total kinetic neutron energy. Intuitively, this is not a surprise, since the total non-em energy consists of other components than just neutrons, and the total kinetic energy of the neutrons is 
not an exact measure for the {\sl number} of neutrons (which is the parameter expected to be correlated to the binding energy loss).
\begin{figure*}[hbt]
\epsfysize=12cm
\centerline{\epsffile{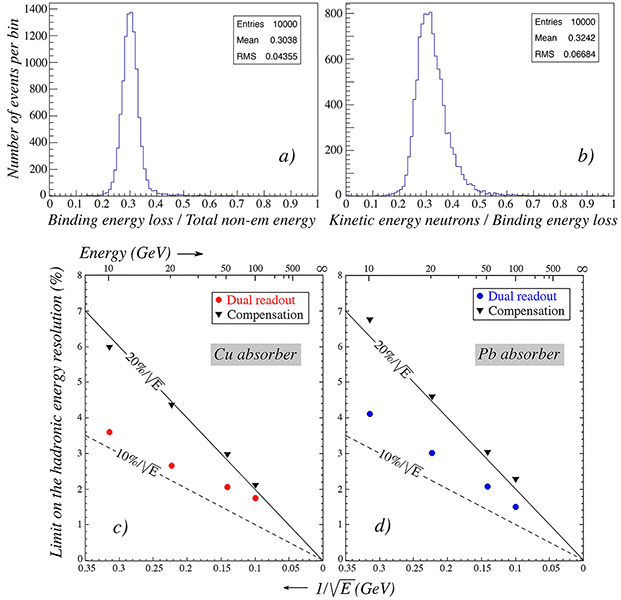}}
\caption{\small
Distributions of the ratio of the non-em energy and the nuclear binding energy loss ($a$) and the ratio of the total kinetic energy carried by neutrons and the nuclear binding energy loss ($b$) for showers generated by 50 GeV $\pi^-$ in a massive block of copper. The limits on the hadronic energy resolution derived from the correlation between the nuclear binding energy losses and the parameters measured in dual-readout or compensating calorimeters, as a function of energy, as a function of the particle energy. The straight lines represent resolutions of $20\%/\sqrt{E}$ and  $10\%/\sqrt{E}$, and are intended for reference purposes. Results from GEANT4 Monte Carlo simulations of pion showers developing in a massive block of copper ($c$) or lead ($d$) \cite{LLW17}.}
\label{sehwook}
\end{figure*}

This explanation was recently tested with dedicated Monte Carlo simulations of hadronic shower development in lead and copper absorber structures \cite{LLW17}.
The simulations were carried out with the GEANT4 Monte Carlo package \cite{geant}.
Events were generated with GEANT4.10.3 patch-02 (released in July 2017). 
For applications of calorimetry in high energy physics, GEANT4 recommends to use the FTFP\_BERT physics list which contains the Fritiof model~\cite{Fritiof}, coupled to the Bertini-style cascade model~\cite{Bertini} and all standard electromagnetic processes. 
This is the default physics list used in simulations for the CMS and ATLAS experiments at CERN's Large Hadron Collider~\cite{g4_pl}.

Pions of different energies were absorbed in these structures. For each event, the following information was extracted: 
\begin{enumerate}
\item The {\sl em shower fraction}, $f_{\rm em}$.  
\item The {\sl total kinetic neutron energy}, $E_{\rm kin}(n)$. 
\item The {\sl total nuclear binding energy loss}, $\Delta B$. 
\end{enumerate}

Simulations were carried out for pions of 10, 20, 50 and 100 GeV. For each run, 10,000 events were generated. These (time consuming) simulations yielded a lot of information. In situations where the results could be compared to experimental data, such as those shown in Figure \ref{femprops}, the agreement was good.
For the present purpose, the main point of interest is the correlation between
the nuclear binding energy loss, which is the main culprit for poor hadronic calorimeter performance, and the variables devised to mitigate the effects of that, \ie the total kinetic neutron energy or $f_{\rm em}$.

Some results of these simulations are shown in Figure \ref{sehwook}. The correlation between the nuclear binding energy loss and these measured variables is illustrated by Figures \ref{sehwook}a and \ref{sehwook}b for dual-readout and compensation, respectively. These simulations clearly favor dual-readout,
at least for the case of 50 GeV pions absorbed in copper. By combining results such as those shown in these correlation plots with the average energy fraction carried by the em shower component, it is also possible to determine a lower limit to the hadronic energy resolution that derives from
fluctuations in the invisible energy. This limit is shown as a function of energy in Figure \ref{sehwook}c and \ref{sehwook}d, for pions developing in a massive block of copper or lead, respectively. 

Experimental data shown in this paper also support the conclusion that the correlation exploited in dual-readout calorimeters provides a more precise measurement of the invisible energy. Figure \ref{Pi100C}c shows that the (\v{C}erenkov) signal from the DREAM
fiber calorimeter is actually a superposition of many rather narrow, Gaussian signal distributions. Each sample in this plot
contains events with (approximately) the same $f_{\rm em}$ value, \ie with the same total non-em energy. 
The dual-readout method combines all these different subsamples and centers them around the correct energy value. The result
is thus a relatively narrow, Gaussian signal distribution with the same central value as for electrons of the same energy.

Figure \ref{neutron1}b shows that the same DREAM (\v{C}erenkov) signal is also a superposition of Gaussian signal distributions of a different type. In this  case, each sample consists of events with (approximately) the same total kinetic neutron energy. The dual-readout method may combine all these different subsamples in the same way as described above.
In doing so, the role of the total non-em energy is taken over by the total kinetic neutron energy, and the method becomes thus very similar to the one used in compensating calorimeters.

A comparison between Figures \ref{Pi100C}c and \ref{neutron1}b shows that the signal distributions from the event samples are clearly wider when the total kinetic neutron energy is chosen to dissect the overall signal. This is consistent with our assessment that dual-readout is a more effective way to reduce the effects of fluctuations in invisible energy on the hadronic energy resolution.

Apart from that, dual-readout offers also several other crucial advantages:
\begin{itemize}
\item Its use is not limited to high-$Z$ absorber materials.
\item The sampling fraction can be chosen as desired.
\item The performance does not depend on detecting the neutrons produced in the absorption process. Therefore, there 
is no need to integrate the calorimeter signals over a large detector volume.
\item The signal integration time can be limited for the same reason. 
\end{itemize}

This is not to say that there is no advantage in detecting the neutrons produced in the shower development. In fact, Figure \ref{neutron2} shows that this may further improve the hadronic calorimeter resolution, since $f_{\rm em}$ and $f_n$ are correlated with the nuclear binding energy losses in different ways, and thus may offer complementary benefits. 

\subsection{Choice of absorber material}

Whereas DREAM used copper as absorber material for the dual-readout fiber calorimeter, RD52 never managed to build an equally large detector using this absorber material,
let alone the larger structure that was envisaged to limit the effects of lateral shower leakage. Hadron measurements were therefore done with a lead-based absorber structure.
While the relatively crude sampling structure of the DREAM calorimeter could be achieved by extruding $4 \times 4$ mm$^2$ tubes with a central hole (a commercially available item with a slightly higher cost than that of the raw material, see Figure \ref{Detector}), the much finer sampling structure was only realized by machining grooves in copper plates (Figure \ref{dreamstructure}). On the other hand, lead structures with the required precision could be extruded relatively easily.

One may wonder why the choice of absorber material is so important. There are at least three reasons for preferring copper over lead:
\begin{enumerate}
\item The $Z$ value. Copper has a much lower $Z$ value than lead, 29 \vs 82. This means, among other things, that the $e/mip$ ratio is very different. The $e/mip$ value is important for the low-energy hadronic signals linearity. Figure \ref{zeuslin2} shows that the compensating ZEUS calorimeter, which used depleted uranium ($Z$ = 92) as absorber material, was found to be non-linear for hadrons with kinetic energies below $\sim$5 GeV. This effect, a gradual increase in the hadronic response by 40\% between 5 and 0.5 GeV, is expected to be three times smaller in a copper-based calorimeter. This calorimeter property is very important for the detection of jets from the hadronic decay of intermediate vector bosons, where
a significant fraction of the energy is carried by soft fragments (Figure \ref{webber}).
\item The $\chi$ value. As illustrated in Figure \ref{dr1}, the $\chi$ parameter is the cotangent of the angle $\theta$, which determines the difference between the scintillation and \v{C}erenkov signals from the dual-readout calorimeter. Depending on the calorimeter structure, this angle $\theta$ can vary between 45$^\circ$ ($\chi = 1$) and 90$^\circ$ ($\chi = 0$). When $\theta = 45^\circ$, all hadronic data points in the scatter plot are located on the diagonal, and therefore the two signals do not provide complementary information. When $\theta = 90^\circ$,
the calorimeter is compensating, the scintillation response is the same for em and hadronic showers, and the \v{C}erenkov signals might be used to reduce the effects of fluctuations in the non-em shower component. The larger $\theta$, \ie the smaller the value of $\chi$, the better  the dual-readout mechanism helps improving the hadronic calorimeter performance. The experimental results indicate optimal values of 0.30 for the copper-based DREAM structure and 0.45 for the lead-based RD52 fiber calorimeter.
\item The density. The absorption of hadron showers is governed by the nuclear interaction length, $\lambda_{\rm int}$. The value of $\lambda_{\rm int}$ for copper (15.1 cm) is smaller than that for lead (17.0 cm). This despite the fact that copper has a smaller density: 8.96
\vs 11.3 g$\cdot$cm$^{-3}$. As a result, a typical hadron calorimeter with a depth of $10 ~\lambda_{\rm int}$ for a $4\pi$ experiment at a particle collider would need to have only half the mass when made out of copper, compared to lead. This is of course an important (engineering) consideration when designing an experiment.
\end{enumerate}

\subsection{Limitations and what to do about these}

The factors that limit the performance of a dual-readout fiber calorimeter of the type developed in RD52 fall into two categories:
\begin{enumerate}
\item Factors that affect the electromagnetic performance
\item Factors that derive from the specific conditions faced in a $4\pi$ experiment at a collider
\end{enumerate}

It has been shown \cite{Akc14b} that the main factor limiting the em energy resolution of the RD52 fiber calorimeter is the \v{C}erenkov light yield.
The total contribution of stochastic fluctuations was measured to be 13.9\%/$\sqrt{E}$, of which only 8.9\%/$\sqrt{E}$ could be attributed to sampling fluctuations, \ie the factor that typically determines the em energy resolution of sampling calorimeters with 
scintillator or liquid argon as active material. This difference is due to the small light yield in the fibers that produce the \v{C}erenkov
signals, $\sim$30 p.e./GeV. There are two straightforward ways in which this light yield can be substantially increased:
\begin{enumerate}
\item By aluminizing the upstream ends of the fibers. This will almost double the yield of isotropically produced light.
It would provide additional advantages, such as an increase of the effective light attenuation length of the fibers, and 
an improved possibility to determine the depth of the light production event by event \cite{Aco91c}. This would benefit the hadronic performance and the particle identification capabilities.
\item By increasing the quantum efficiency of the detectors that convert this light into electric signals. In the PMTs used in the RD52 fiber calorimeter, this quantum efficiency was $\sim$20\% for the (predominantly blue) light that constituted the \v{C}erenkov signals.
This value could at least be doubled, and possibly tripled, if silicon-based light detectors were used, such as SiPMs. This was recently experimentally confirmed, when RD52 tested SiPM readout on a small lateral segment of their copper-fiber calorimeter \cite{SPSC17}.
\end{enumerate}

SiPM readout offers also important 
potential advantages for application of dual-readout calorimeters in modern experiments at colliding beam machines:
\begin{itemize}
\item It offers the possibility to eliminate the forests of optical fibers that stick out at the rear end (Figure \ref{detector2}a). These fiber bunches occupy precious space and act as antennas for particles that come from sources other than the showers developing in the calorimeter. 
\item This compact readout makes it possible to separate the calorimeter into longitudinal segments, if so desired \footnote{For a detailed discussion about the implications of longitudinal segmentation of a sampling calorimeter, the reader is referred to chapter 6 of {\em Calorimetry - Energy Measurement in Particle Physics}, International Series of Monographs on Physics, volume 168, Oxford University Press (2017).}.
\item Unlike the PMTs used until now, SiPMs can operate in a magnetic field.
\end{itemize}
There are of course also potential disadvantages, most notably the fact that SiPMs are {\sl digital} detectors and therefore prone to signal saturation effects. A major challenge for this particular calorimeter concerns the fact that the SiPMs have to read the signals from a grid of closely spaced fibers of two different kinds, where the light yield of one type of fibers (detecting the \v{C}erenkov
light) is an order of magnitude smaller than that of the other fibers (detecting the scintillation light). Crosstalk is thus a major concern.

\subsection{Outlook}

The work described in this paper was never more than a generic R\&D effort, intended to investigate the properties of what seemed (20 years ago \cite{Tucson}) a promising new avenue. This research has led to a series of remarkable results:
\begin{itemize}
\item Excellent energy resolution, both for em and hadron showers
\item Correct reconstruction of the energy of all particles that are absorbed in the detector
\item No difference in the response to protons, pions and kaons
\item The possibility to measure the ionization and radiation energy losses by muons traversing the calorimeter {\sl separately}
\item Excellent particle ID capabilities, including the recognition and identification of electrons that are part of collimated jets
\item All of the above can be achieved in an instrument calibrated with electrons.
\end{itemize}
%

If one would want to base a future collider experiment on the dual-readout calorimeter technology, then a number of practical challenges will have to be dealt with. 
Any follow-up of the RD52 project will have to address these challenges, and develop acceptable solutions. The SiPM work mentioned above is a first step in this process.  
Important other challenges concern the large-scale production of an absorber structure that has to meet very tight specifications, and the issues deriving from the need to make the detector structure projective. it is very non-trivial to make a 4$\pi$ detector structure with longitudinal optical fibers, although some useful ideas have been pursued with that purpose in mind \cite{Anz95b}. Attempts by the RD52 Collaboration to find practical solutions for these challenges have yielded insufficient results, primarily because of a lack of resources. However, this would probably change if a commitment were made to use this new detector concept for a large new experiment.

At present, experiments planned for the proposed high-energy electron-positron colliders FCCee (CERN) and CEPC (China) are seriously considering calorimeter systems based on the dual-readout technique. Experiments at future high-energy hadron colliders (including LHC upgrades) are expected to benefit much less from the advantages offered by calorimeters such as those described in this paper. The performance of calorimeters in such experiments is likely to be dominated by the extremely high event rates needed to extract new physics results, and by the effects of the Lorentz-boosted center-of-mass of the fundamental (constituent) collisions. The main benefit for experiments at hadron colliders might actually be the instantaneous character of the \v{C}erenkov radiation, and the associated promise of the ultrafast (10 ps) timing needed to distinguish events occurring in the same bunch crossing.

\section*{Acknowledgments}

We would like to thank all our wonderful colleagues from the ACCESS, DREAM and RD52 Collaborations, who have contributed to the work that led to the many experimental results shown in this review. That includes not only the scientists and students from the participating institutions, but also the engineers and technicians who built the instruments that were tested. We would also like to express our appreciation for the help received from CERN, which provided the test beams and the related infrastructure that made these tests possible.
Financial support for this research has been received from the United States Department of Energy, under contract DE-FG02-12ER41783, from Italy's Istituto Nazionale di Fisica Nucleare and Ministero dell' Istruzione, dell' Universit\`a e della Ricerca, and from the Basic Science Research Program of the National Research Foundation of Korea (NRF), funded by the Ministry of Science, ICT \& Future Planning under contract 2015R1C1A1A02036477.

\renewcommand\bibname{References}

\end{document}